\pdfoutput=1
\documentclass[%
 reprint,
superscriptaddress,
 amsmath,amssymb,
 aps,
 pra,
]{revtex4-2}
\usepackage{caption}
\usepackage{subcaption}
\usepackage{graphicx}
\usepackage{amsmath}
\usepackage{geometry}
\usepackage{hyperref}
\usepackage[T1]{fontenc}
\usepackage[utf8]{inputenc}
\usepackage{float}
\usepackage{csquotes}
\usepackage[version=4]{mhchem}
\usepackage{amssymb}
\usepackage{xcolor}
\usepackage{bbold}
\usepackage{dcolumn}
\usepackage{bm}
\usepackage{algorithm}
\usepackage{algpseudocode}

\usepackage{tikz}
\usetikzlibrary{quantikz2}

\begin{document}

\preprint{}

\title{
A Feasible Design of Elementary Quantum Arithmetic Logic Units for Near-Term Quantum Computers
}

\author{Junxu Li}
\email{Emial: lijunxu1996@gmail.com}
\affiliation{Department of Physics, College of Science, Northeastern University, Shenyang 110819, China}

\date{\today}

\begin{abstract}
Quantum arithmetic logic units (QALUs) constitute a fundamental component of quantum computing. 
However, the implementation of QALUs on near-term quantum computers remains a substantial challenge, largely due to the limited connectivity of qubits. 
In this paper, we propose feasible QALUs, including quantum binary adders, subtractors, multipliers, and dividers, which are designed for near-term quantum computers with qubits arranged in two-dimensional arrays. 
Additionally, we introduce a feasible quantum arithmetic operation to compute the two's complement representation of signed integers.
The proposed QALUs utilize only Pauli-X gates, CNOT gates, and $C\sqrt{X}$ (CSX) gates, and all two-qubit gates are operated between nearest neighbor qubits. 
Our work demonstrates a viable implementation of QALUs on near-term quantum computers, advancing towards scalable and resource-efficient quantum arithmetic operations.
\end{abstract}

\maketitle

\section{Introduction}

Quantum computing represents a paradigm shift in computational capabilities.
The unitary quantum operations affect simultaneously each element of the superposition, facilitating extensive parallel data processing within a singular quantum hardware\cite{deutsch1985quantum}.
By leveraging superposition and entanglement, quantum computers usher in unprecedented computational capabilities, and can efficiently solve some classically intractable problems\cite{deutsch1992rapid, bernstein1993quantum, lloyd1996universal, ekert1996quantum, simon2011quantum, rebentrost2014quantum, preiss2015strongly, biamonte2017quantum, preskill2018quantum, li2019quantum, cao2019quantum, google2020hartree, huang2021power, altman2021quantum, schlimgen2021quantum, sajjan2022quantum, ma2023multiscale}.
The best known example is Shor's algorithm\cite{shor1994algorithms}, which is designed to find the prime factors of a large composite integer.
Besides, quantum states that are easily prepared with a quantum computer exhibit properties beyond classical capabilities\cite{lund2017quantum, harrow2017quantum, arute2019quantum, zhong2020quantum}, and no known classical algorithm has been able to simulate a quantum computer to date\cite{preskill2018quantum}.

At the core of this transformative technology are quantum arithmetic logic units (QALUs), which perform essential arithmetic operations such as addition, subtraction, multiplication, division, and modular exponentiation  within a quantum framework\cite{vedral1996quantum, beckman1996efficient}.
Classical computing relies on arithmetic logic units (ALUs) to execute basic arithmetic operations\cite{burks1946preliminary, horowitz1989art}.
Similarly, QALUs serve as the building blocks for more complex computations and algorithms in quantum computing.
For example, the quantum modular arithmetic is integral to the Shor's factoring algorithm\cite{shor1994algorithms}.
In this context, the efficiency of QALUs is directly correlated with the overall performance of the quantum algorithm.
The creation of efficient and practical QALUs is essential for unlocking the full potential of quantum computing, paving the way for efficient quantum algorithms and applications\cite{takahashi2009quantum, ruiz2017quantum}.
Research on quantum arithmetic circuits is not only of theoretical significance but also poised to drive technological breakthroughs and innovative applications. 

Thus, there is a broad range of studies in the existing literature which tackle the design of QALUs for quantum systems, and it is still of great interest in developing novel designs and refining the existing ones.

A primary design method of QALUs utilizes the Clifford gates and $T$ gates sets\cite{wang2024comprehensive}.
The first QALUs were reversible versions of known classical implementations\cite{vedral1996quantum, beckman1996efficient}.
In 1996, Vedral and coworkers explicitly constructed several quantum elementary arithmetic operations using Clifford gates and $T$ gates\cite{vedral1996quantum}, including the first quantum ripple-carry adder and the quantum modular exponentiation, which is a significant landmark in the development of quantum arithmetic.
To perform binary addition on quantum computers, researchers have proposed and developed various quantum adders based on ripple-carry\cite{vedral1996quantum, cuccaro2004new,gidney2018halving}, carry-lookahead\cite{draper2004logarithmic, wang2023reducing}, carry-save\cite{gossett1998quantum}, and hybrid\cite{takahashi2008fast,takahashi2009quantum} structures.
The quantum binary subtractors have structures closely resembling those of quantum adders, and the typical quantum subtractors are implemented based on ripple-borrow subtraction method\cite{cheng2002quantum, thapliyal2009design, thapliyal2016mapping}.
In the domain of multiplication, various Clifford and $T$ gates based quantum multipliers are proposed, leveraging the typical multiplication methods including long multiplication\cite{lin2014qlib, parent2017improved, munoz2018quantum}, Karatsuba\cite{kepley2015quantum, gidney2019asymptotically}, Toom-Cook\cite{dutta2018quantum, putranto2023space}, and Wallace Tree\cite{gayathri2021t, orts2023improving} algorithms.
Recently, inspired by classical division algorithms, such as long division and Goldschmidt algorithm, several quantum dividers are also presented using Clifford and $T$ gates\cite{thapliyal2019quantum, gayathri2021t, yuan2022novel, orts2024quantum}.
In these QALUs, Toffoli gates play a vital role.
To standardize and compare various quantum designs, the $T$ gate cost is frequently discussed, particularly at the level of the Toffoli gate\cite{jang2023quantum}.
While Clifford gates are relatively straightforward to implement and exhibit lower error rates, $T$ gates are more complex and costly, making their optimization crucial for the efficient design of quantum circuits.
Circuits designed with this method can exploit advanced fault-tolerant architectures based on quantum error-correcting codes, such as the surface code\cite{google2023suppressing}, to address critical challenges such as the fragility of quantum states and susceptibility to external noise.

Another significant arithmetic framework is based on the quantum Fourier transform (QFT)\cite{weinstein2001implementation, coppersmith2002approximate}.
QFT-based quantum arithmetic circuits typically commence with a QFT block that converts the input state to the frequency domain, performs arithmetic operations using controlled phase gates, and then reverts the state to the original domain through an inverse Fourier transform.
With assistance of the QFT, a variety of fast integer arithmetic operations are developed, including QFT-based quantum adders\cite{draper2000addition, cheng2002quantum}, subtractors\cite{beauregard2002circuit, csahin2020quantum}, multipliers\cite{beauregard2002circuit, maynard2014quantum} and dividers\cite{khosropour2011quantum}.
This approach leverages efficient operations in the Fourier basis to streamline arithmetic operations, particularly in scenarios requiring multiplication and exponentiation\cite{ruiz2017quantum}.

Despite advancements in quantum computing, the implementation of QALUs on near-term quantum computers remains challenging
One inevitable obstacle is the implementation of Toffoli gates, which are crucial in QALUs based on Clifford and T gates. 
Experimentally, achieving high-fidelity Toffoli gates is exceedingly difficult, even for a linear chain of three qubits\cite{kim2022high}.
A Toffoli gate can be decomposed into single- and two-qubit gates. 
The decomposition requires at least five two-qubit gates for fully connected qubits\cite{barenco1995elementary, yu2013five}, and eight for nearest-neighbor connected qubits\cite{smith2023leap}.
The complexity increases further when Toffoli gates involve qubits that are not physically connected\cite{li2023toward, li2023m}.
Another obstacle is the implementation of two- or multi-qubit gates among the remote qubits\cite{siraichi2018qubit, cowtan2019qubit}.
In existing QALUs, there are often two-qubit or multi-qubit gates that involve qubits which are not physically connected.
In many quantum computing architectures, qubits are arranged in physical layouts where not all qubits are directly connected\cite{linke2017experimental, arute2019quantum, zhu2022quantum}.
Consequently, two-qubit gates are typically executed on qubits that are physically adjacent.
The limited connectivity of native qubits poses a significant hindrance to the implementation of quantum algorithms that necessitate long-range interactions.

To address these issues, this paper presents elementary quantum arithmetic logic units (QALUs), including quantum adders, subtractors, multipliers, and dividers, designed for near-term quantum computers.
The proposed QALUs can be implemented using only Pauli-X gates, CNOT gates, and $C\sqrt{X}$ gates,with all two-qubit gates operating on neighboring qubits.
Thus, the proposed QALUs are feasible for implementation on near-term quantum computers where qubits are organized in two-dimensional arrays, such as Sycamore\cite{arute2019quantum} and Zuchongzhi\cite{zhu2022quantum}. 

The rest of the paper is organized as follows.
In Sec.(\ref{sec_adder}), we introduce scalable quantum full adders (denoted as $P_I$, $P_{II}$), where input terms are mapped to two qubit columns and the output sum is mapped to another column. 
Additionally, we present a special quantum adder (denoted as $P_{III}$) optimized for iterative additions. 
For subtraction, in Sec.(\ref{sec_subtractor_main}), we present the quantum operation that compute the two's complement representation of signed binary integers, and demonstrate how to perform subtraction with quantum adders in the  two's complement representation.
Inspired by the principles of long multiplication and long division, we then present quantum multipliers and quantum dividers in Sec.({\ref{alg_qmul}}) and Sec.(\ref{sec_divider}).
For simplicity, in Sec.(\ref{sec_adder}, \ref{sec_subtractor_main}, {\ref{alg_qmul}}, \ref{sec_divider}) we assume that the input states are some certain states in the computational basis.
In Sec.(\ref{sec_discussion}),  we explicitly demonstrate that the proposed QALUs also support superposition states as input.
Our conclusions are presented in Sec.(\ref{sec_conclusion}).

\noindent

\section{Quantum Adders}
\label{sec_adder}

\subsection{Quantum One-bit Full Adder}
\label{Quantum One bit Full Adder}
Addition is a fundamental operation that serves as a cornerstone for the functionality of all algorithms built upon it.
In classical electronics, adders, or summers, are digital circuits that perform addition of binary numbers\cite{burks1946preliminary, horowitz1989art}.
In 1937, Claude Shannon firstly demonstrated binary addition in his graduate thesis\cite{shannon1938symbolic}.
One-bit full adder adds three inputs and produces two outputs.
The first two inputs are single binary digits $a$ and $b$, and the third input is an input carry, denoted as $c_{in}$.
The carry digit represents an overflow into the next digit of a multi-digit addition.
The two outputs are sum $s$ and carry $c_{out}$, where $a + b + c_{in} = s + 2c_{out}$.
A full adder performs binary addition as
\begin{equation}
    s = a \oplus b \oplus c_{in}
    \label{eq_adder_s}
\end{equation}
\begin{equation}
    c_{out} = (a\cdot b)+(c_{in}\cdot(a\oplus b))
    \label{eq_adder_cout}
\end{equation}

\begin{figure}[t]
    \centering
    \includegraphics[width=0.4\textwidth]{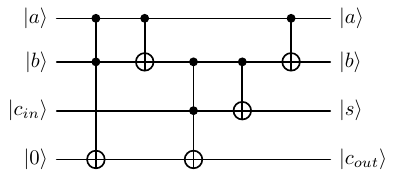}
    \caption{One typical implementation of a one-bit full adder with Toffoli gates and CNOT gates.}
    There are 4 qubits, 2 qubits for the binary digits $a$ and $b$, and the other 2 qubits for the carry and the sum.
    The input $c_{in}$ is converts into $s$ as output, whereas $c_{out}$ takes the place of $c_{in}$ in the next digit of a multi-digit addition.
    \label{qadder_1}
\end{figure}

Traditionally, addition algorithms designed for a quantum computer have mirrored their classical counterparts\cite{beckman1996efficient, vedral1996quantum}. 
One possible implementation of quantum one-bit full adder is depicted in Fig.(\ref{qadder_1}), which is implemented with only Toffoli gates and CNOT gates.
There are 4 qubits, 2 qubits for the binary digits $a$ and $b$, and the other 2 qubits for the carry and the sum.
The input $c_{in}$ is converts into $s$ as output, whereas $c_{out}$ takes the place of $c_{in}$ in the next digit of a multi-digit addition.

In the quantum adder as shown in Fig.(\ref{qadder_1}), there are two Toffoli gates.
However, it is experimentally challenging to implement high-fidelity Toffoli gates, and it is hardly feasible to implement multi-digit addition based on this design on near-term quantum computers. 

To address this issue, here we present a feasible and scalable quantum full-adder, which can be decomposed into Pauli-X gates, $C\sqrt{X}$ gates and CNOT gates, and all two-qubit gates are only applied on the nearest-neighbor qubits.
Therefore, it is feasible to implement the proposed quantum adder on near-term quantum computers, where the qubits are assigned in two-dimensional arrays.

A quantum one-bit full adder performs the binary addition as shown in Eq.(\ref{eq_adder_s}) and Eq.(\ref{eq_adder_cout}), which can be rewritten as,
\begin{equation}
    |s\rangle = X^{a\oplus b}|c_{in}\rangle
    \label{eq_qadder_s}
\end{equation}
\begin{equation}
    |c_{out}\rangle
    =
    X^{a\oplus b\oplus c_{in} + 
    \frac{1}{2}(a\oplus b + a\oplus c_{in} +b\oplus c_{in})}
    |0\rangle
    \label{eq_qadder_c}
\end{equation}
where $X$ is the Pauli-X gate, and $X^{1/2}$, often denoted as $\sqrt{X}$ or $SX$ gate, is the square root of $X$,
\begin{equation}
    \sqrt{X} = \frac{1}{2}
    \begin{bmatrix}
        1+i &1-i
        \\
        1-i &1+i
    \end{bmatrix}
\end{equation}

According to Eq.(\ref{eq_qadder_s},\ref{eq_qadder_c}), there are at least 4 qubits in a quantum one-bit full adder.
Two qubits denoted as $A$, $B$ for the inputs $a$, $b$, one qubit denoted as $C$ for the input carry $c_{in}$ and output $s$, and one qubit denoted as $C'$ for the output carry $c_{out}$.
States of qubits $A$, $B$ remains unchanged in the addition operation.

\begin{figure}[t]
    \centering
    \includegraphics[width=0.4\textwidth]{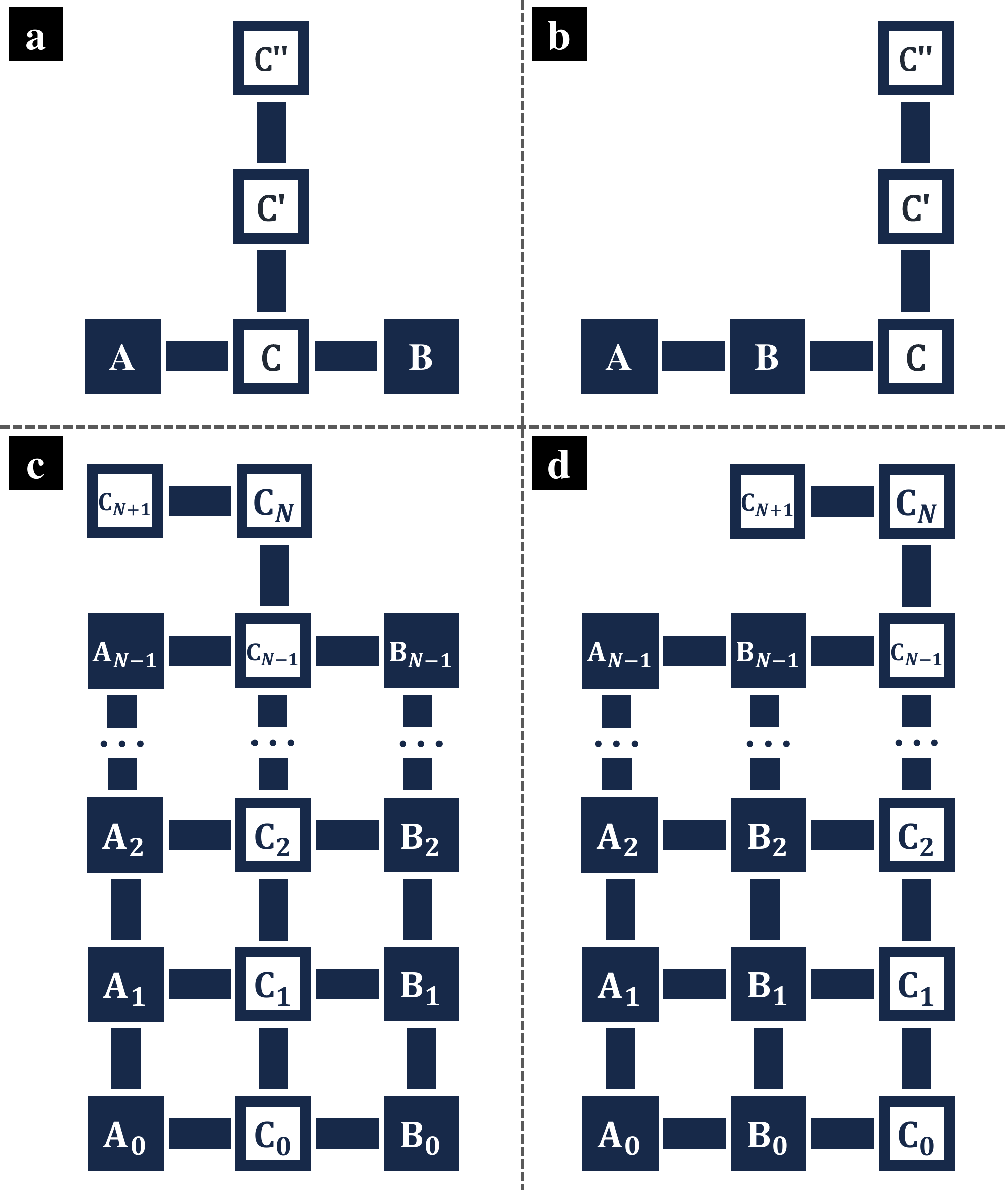}
    \caption{
    {\bf Required qubit and their connectivity for the proposed quantum adders.}
    Squares in the graph represent physical qubits and edges are the allowed interactions.
    (a, b) Qubit connectivity for 2-bit quantum full adder.
    Inputs $a$, $b$ are mapped onto qubits $A$, $B$, $c_{in}$ is mapped onto qubit $C$.
    Qubits $C'$, $C''$ are initialized at state $|0\rangle$.
    After the applying the circuits of quantum full adder, the state of qubit $C$ indicates the sum $s$, whereas state of $C'$ indicates the output carry $c_{out}$.
    $C''$ is ancilla qubit.
    (c, d) Quantum adders supporting multiple bits, which are cascades of one-bit full adders as shown in (a) and (b).
    In (a, c), the outputs are at stored in the middle column, whereas in (b, d) the outputs are stored in the side column.
    }
    \label{fig_adder_qubit}
\end{figure}

Here we present two designs of quantum one-bit one bit full adders, and in Fig.(\ref{fig_adder_qubit}a,b) we present the corresponding qubit connectivity, where squares represent physical qubits and edges are the allowed interactions.
A cascade of these quantum one-bit full adders forms quantum adders for multiple bits, and the corresponding qubit connectivity is depicted in Fig.(\ref{fig_adder_qubit}c,d).
Details about quantum adders for multiple bits are later discussed in Sec.(\ref{Quantum Adders Supporting Multiple Bits}).
In Fig.(\ref{fig_adder_qubit}a,c) the two inputs are not physically connected, and the outputs are stored in the center column.
In Fig.(\ref{fig_adder_qubit}b,d) the two inputs are connected, whereas the outputs are stored in the side column.
In Fig.(\ref{fig_adder_qubit}a,b), qubit $C''$ is ancilla qubit, which is initialized at state $|0\rangle$ and is still at state $|0\rangle$ after the full addition operation.
For simplicity, we denote the one-bit full adders in Fig.(\ref{fig_adder_qubit}a,b) as $\hat{P}_{I}$, $\hat{P}_{II}$, and the corresponding adders for multiple bits as $P_I$, $P_{II}$.


\begin{figure*}[ht]
    \centering
    \includegraphics[width=0.9\textwidth]{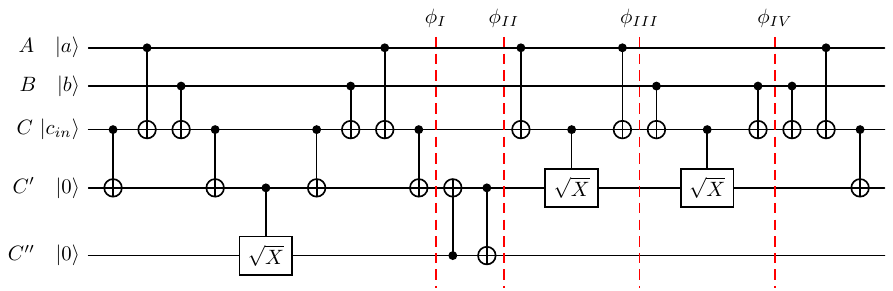}
    \caption{Implementation of quantum one-bit full adder $\hat{P}_{I}$, corresponding to the qubit connectivity in Fig.(\ref{fig_adder_qubit})a.
    There are 5 qubits, 2 qubits denoted as $A$, $B$ for the binary digits $a$ and $b$, 2 qubits $C$ and $C'$ for the carry and the sum, and qubit $C''$ as ancilla qubit.
    Initially, inputs $a$, $b$ are mapped to qubits $A$, $B$, and the input carry $c_{in}$ is mapped to qubit $C$.
    Qubits $C'$ and $C''$ are initialized at state $|0\rangle$. }
    \label{fig_qadder_1}
\end{figure*}

Consider the situation where the outputs in the middle columns as depicted in Fig.(\ref{fig_adder_qubit}a).
There are 5 qubits denoted as $A,B,C,C',C''$ , and the edges are the allowed interactions between the nearest-neighbor qubits.
In Fig.(\ref{fig_qadder_1}) we present the corresponding quantum one-bit full adder.
There are in total 20 two-qubit gates, including 17 CNOT gates and 3 $C\sqrt{X}$ gates.
All of these two-qubit gates are applied between  nearest-neighbor qubits as depicted in Fig.(\ref{fig_adder_qubit}a).
Thus, $\hat{P}_{I}$ can be implemented on near-term quantum computers, where the qubits are assigned in two-dimensional arrays.

In Tab.(\ref{tab_qadder}) we present the quantum states evolution when applying $\hat{P}_I$, where $\phi$ corresponds to the stages marked by the dashed lines in Fig.(\ref{fig_qadder_1}).
Initially, inputs $a$, $b$ are mapped to qubits $A$, $B$, and the input carry $c_{in}$ is mapped to qubit $C$.
Qubits $C'$ and $C''$ are initialized at state $|0\rangle$.
For simplicity, the overall quantum states are written as a separable form as shown in Tab.(\ref{tab_qadder}), where the inputs $a,b,c_{in}\in\{0,1\}$.
The first step is to implement operation $X^{\frac{1}{2}a\oplus b}$, corresponding to the operations before $|\phi_I\rangle$ in Fig.(\ref{fig_qadder_1}).
Recalling the qubit connectivity as shown in Fig.(\ref{fig_adder_qubit}a), qubits $A$ and $B$ are not physically connected.
Notice that $c_{in}\oplus(a\oplus b\oplus c_{in})=a\oplus b$, here we alternatively prepare qubit $C'$ at state $|a\oplus b\rangle$ by the first 4 CNOT gates.
Next, a $C\sqrt{X}$ gate is applied, where $C'$ is control qubit and $C''$ is the target.
Then the 4 CNOT gates are applied in the inverse order, converting $C'$ and $C''$ to the initial state.
The second step is to swap the state $X^{\frac{1}{2}a\oplus b}|0\rangle$ to qubit $C'$, according to the two CNOT gates between $\phi_I$ and $\phi_{II}$.
Generally, a SWAP gate contains three CNOT gates.
Here the initial state of $C'$ is $|0\rangle$, thus we only need two CNOT gates to swap the state of qubits $C'$ and $C''$.
Since then we will add no more operations on ancilla qubit $C''$.
Ancilla qubit $C''$ keeps state $|0\rangle$ till the end.
The third step is to implement operation $X^{\frac{1}{2}a\oplus c_{in}}$, corresponding to the operations between $\phi_{II}$ and $\phi_{III}$.
Similarly, the forth step is to implement operation $X^{\frac{1}{2}b\oplus c_{in}}$, corresponding to the operations between $\phi_{III}$ and $\phi_{IV}$.
The final step is to prepare the $a\oplus b\oplus c_{in}$ terms, corresponding to the three CNOT gates after $\phi_{IV}$.
The first two CNOT gates converts $C$ from $|c_{in}\rangle$ to $|a\oplus b\oplus c_{in}\rangle$, and the last CNOT gate implements $X^{a\oplus b\oplus c_{in}}$ on qubit $C'$.
As $X$ and $\sqrt{X}$ gates commute, the output state of $C$ and $C'$ are the sum $s$ and output carry $c_out$ in two-bit binary addition, as described in Eq.(\ref{eq_qadder_s}, \ref{eq_qadder_c}).

The quantum one-bit full adder in Fig.(\ref{fig_qadder_1}) is denoted as $\hat{P}_{I}(A, B, C, C', C'')$, where $P$ indicates that the circuit is designed for addition operation (plus), subscript $I$ indicates that this is the first type quantum adder, the hat notation indicates that this is one-bit quantum adder.
For clarity, the involved qubits, $A, B, C, C', C''$, are also included in the notation.
The computational basis can be expressed as a five-digit binary string, which corresponds to these five qubits.
We have
\begin{equation}
    \hat{P}_{I}(A, B, C, C', C'')
    |a,b,c_{in},0,0\rangle
    =|a,b,s,c_{out},0\rangle
    \label{eq_p1_I}
\end{equation}
where the outputs $s$, $c_{out}$ are given in Eq.(\ref{eq_adder_s},\ref{eq_adder_cout}).

\begin{table*}
\caption{\label{tab_qadder}Overall quantum states at certain stages in the quantum one-bit full adder, $\phi$ corresponds to the dashed lines marked in Fig.(\ref{fig_qadder_1}) and Fig.(\ref{fig_qadder_2}).}
\begin{ruledtabular}
\begin{tabular}{ cccccc  }
 \multicolumn{6}{c}{Overall quantum state} \\
 \hline
 Stage &\multicolumn{5}{c}{Qubit} \\
  &A &B &C &C' &C''\\
 \hline
 Initial &$|a\rangle$ &$|b\rangle$ &$|c_{in}\rangle$ &$|0\rangle$ &$|0\rangle$ \\
 $\phi_{I}$ &$|a\rangle$ &$|b\rangle$ &$|c_{in}\rangle$ &$|0\rangle$ &$X^{\frac{1}{2}a\oplus b}|0\rangle$\\
 $\phi_{II}$ &$|a\rangle$ &$|b\rangle$ &$|c_{in}\rangle$ &$X^{\frac{1}{2}a\oplus b}|0\rangle$ &$|0\rangle$\\
 $\phi_{III}$ &$|a\rangle$ &$|b\rangle$ &$|c_{in}\rangle$ &$X^{\frac{1}{2}(a\oplus b+a\oplus c_{in})}|0\rangle$ &$|0\rangle$\\
 $\phi_{IV}$ &$|a\rangle$ &$|b\rangle$ &$|c_{in}\rangle$ &$X^{\frac{1}{2}(a\oplus b+a\oplus c_{in}+b\oplus c_{in})}|0\rangle$ &$|0\rangle$\\
 Final &$|a\rangle$ &$|b\rangle$ &$|a\oplus b\oplus c_{in}\rangle$ &$X^{a\oplus b\oplus c_{in} + \frac{1}{2}(a\oplus b+a\oplus c_{in}+b\oplus c_{in})}|0\rangle$ &$|0\rangle$\\
\end{tabular}
\end{ruledtabular}
\end{table*}

\begin{figure*}[th]
    \centering
    \includegraphics[width=0.9\textwidth]{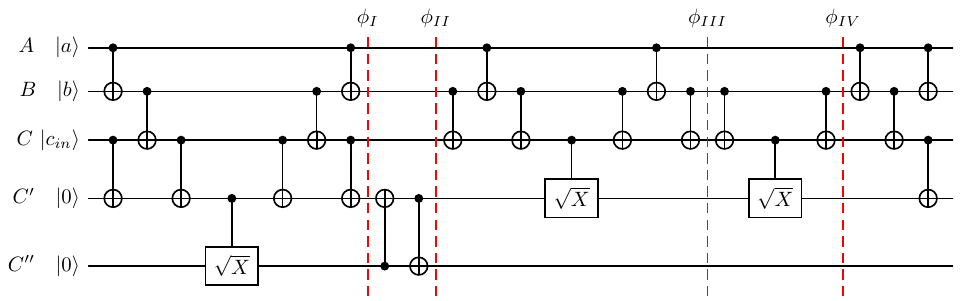}
    \caption{Implementation of quantum one-bit full adder $\hat{P}_{II}$, corresponding to the qubit connectivity in Fig.(\ref{fig_adder_qubit}b).
    The quantum states at certain stages are exactly the same as presented in Tab.(\ref{tab_qadder}), where $\phi$ indicate the stages marked by the dashed lines.}
    \label{fig_qadder_2}
\end{figure*}

In the quantum one-bit full adder as shown in Fig.(\ref{fig_qadder_1}), qubits $A$, $B$ that represent the inputs are not physically connected, and the outputs are stored in the middle column.
The qubits might be assigned in other architectures.
Consider the situation as depicted in Fig.(\ref{fig_adder_qubit}b), where the qubits $A$, $B$ are neighbors, and only $B$ directly connects to qubit $C$, which is initialized as the input carry $|c_{in}\rangle$.  
Similarly, there are 5 qubits denoted as $A,B,C,C',C''$, and the edges in Fig.(\ref{fig_adder_qubit}b) represent the allowed interactions between the neighbors.
The quantum states at certain stages are exactly the same as presented in Tab.(\ref{tab_qadder}), where $\phi$ indicate the stages marked by the dashed lines in Fig.(\ref{fig_qadder_2}).

We denote the quantum one-bit full adder as shown in Fig.(\ref{fig_qadder_2}) as $\hat{P}_{II}(A, B, C, C', C'')$, indicating the second type quantum adder.
Similarly, we have
\begin{equation}
    \hat{P}_{II}(A, B, C, C', C'')|a,b,c_{in},0,0\rangle
    =|a,b,s,c_{out},0\rangle
    \label{eq_p1_II}
\end{equation}
As qubit $A$ is not directly connected to qubit $C$, here  additional CNOT gates are necessary to implement the operations such as $X^{\frac{1}{2}a\oplus c_{in}}$.
Thus, the quantum full-adder in Fig.(\ref{fig_qadder_2}) requires more CNOT gates.
In total, the quantum one-bit full-adder in Fig.(\ref{fig_qadder_2}) can be decomposed into 22 CNOT gates and 3 $C\sqrt{X}$ gates, whereas the implementation in Fig.(\ref{fig_qadder_1}) requires 17 CNOT gates and 3 $C\sqrt{X}$ gates.
Notice that there are repeating CNOT pairs at the left and right of the dashed line $\phi_{IV}$ in Fig.(\ref{fig_qadder_1}), and at the left and right of $\phi_{III}$ in Fig.(\ref{fig_qadder_2}).
These CNOT pairs cancel out and can be deleted in the final implementation.
Therefore, the proposed quantum one-bit full-adder requires 15 CNOT gates and 3 $C\sqrt{X}$ gates with the implementation in Fig.(\ref{fig_qadder_1}), or 20 CNOT gates and 3 $C\sqrt{X}$ gates with the implementation in Fig.(\ref{fig_qadder_2}).

\subsection{Standard Quantum Adders Supporting Multiple Bits}
\label{Quantum Adders Supporting Multiple Bits}
In classical computers or processors, the one-bit full adder is usually a component in a cascade of adders that adds multiple bit numbers.
Similarly, it is possible to create a quantum adder that supports multiple bits by cascading multiple one-bit full adders.
In this section, we propose the design of quantum adder that adds two given integers.
The inputs are two N-bit binary integers, and the output is a (N+1)-bit binary integer.
The inputs are mapped to $2N$ qubits, and the output is stored in $N+1$ qubits.
Additionally, one extra ancilla qubit is involved in the quantum adder.
Thus, $3N+2$ qubits are required in total.

Consider $3N+2$ qubits as depicted in Fig.(\ref{fig_adder_qubit}c,d), where the squares represent physical qubits and edges are the allowed interactions.
The $N$ qubits denoted as $A_{N-1},\cdots,A_{0}$ are initialized as $|a_{N-1}\cdots a_0\rangle$, corresponding to the input integer in N-digit binary form $(a_{N-1}\cdots a_0)_{bin}$, where $a_{N-1}$ is the highest digit and $a_0$ is the lowest digit.
Similarly, the $N$ qubits denoted as $B_{N-1},\cdots,B_{0}$ are initialized as $|b_{N-1}\cdots b_0\rangle$, corresponding to the input integer in N-digit binary form $(b_{N-1}\cdots b_0)_{bin}$.
The $N+2$ qubits qubits denoted as $C_{N+1},\cdots,C_{0}$ are all initialized as ground state $|0\rangle$.

The quantum multi-bit adder can be implemented as a chain of cascaded quantum one-bit full adders.
The first step is the addition of the lowest digit.
5 qubits are included in the one-bit addition: $A_0, B_0, C_0, C_1, C_2$.
Qubits $A_0$ and $B_0$ corresponds to the inputs $a_0$ and $b_0$, which are the least significant bits (LSB) of the inputs.
The input carry $c^0_{in}$ for the addition of lowest digit is 0, thus qubit $C_0$ is initialized as $|0\rangle$ at the beginning.
Qubits $C_1$, $C_2$ are also initialized as $|0\rangle$.
The superscripts of the input carry and output carry indicate the corresponding order of digit.
For instance, in the addition of the LSB, the input carry is denoted as $c^0_{in}$, and the output carry is denoted as $c^0_{out}$.
To add the LSB, apply the quantum one-bit full adder $\hat{P}_{I}$ or $\hat{P}_{II}$ as presented in Sec.(\ref{Quantum One bit Full Adder}).
Qubits $A_0$, $B_0$ do not change in the addition.
Qubit $C_0$ is converted to state $|s_0\rangle$, where $s_0$ is the output sum of $a_0$ and $b_0$.
Qubit $C_1$ is converted to $|c_{out}^0\rangle$, corresponds to the output carry in addition of $a_0$ and $b_0$.
The ancilla qubit $C_2$ is still at state $|0\rangle$.

Then consider the addition of the next digit, $a_1$, $b_1$, along with input carry $c^1_{in}$.
In multi-bit addition, the output carry in the addition of $n^{th}$ digit is the input carry in the addition of $(n+1)^{th}$ digit.
Here qubit $C_1$ is converted to $|c_{out}^0\rangle$ after the addition of $a_0$, $b_0$, where $c_{out}^0$ is also the input carry $c^1_{in}$ for the addition of $a_1$, $b_1$.
Thus after the addition of $a_0$, $b_0$, we have qubits $A_1$, $B_1$, $C_1$ at the appropriate state corresponding to the inputs for the addition of $a_1$, $b_1$.
Apply the quantum one-bit full adder  $\hat{P}_{I}$ or $\hat{P}_{II}$  on qubits $A_1, B_1, C_1, C_2, C_3$.
Qubit $C_1$ is converted to state $|s_1\rangle$, where $s_1$ is the output sum of $a_1$ and $b_1$.
Qubit $C_2$ is converted to $|c_{out}^1\rangle$, corresponds to the output carry in addition of $a_1$ and $b_1$.
The ancilla qubit $C_3$ is still at state $|0\rangle$.

By iteratively adding each bit, we apply the quantum one-bit full adder $N-2$ times, from the less significant bit to the more significant ones.
By the end, after the addition of the most significant bit (MSB), qubit $C_N$ is converted to state $|c_{out}^{N-1}\rangle$, corresponding to the output carry in addition of $a_{N-1}$ and $b_{N-1}$.
There is already no more higher order digits in the inputs.
Thus, the output carry $c_{out}^{N-1}$ is just the MSB in the input sum, and denoted as $s_{N}$ for simplicity.
Therefore, the state of qubits $C_N,\cdots,C_0$ corresponds to the output, a $(N+1)$-bit binary integer.
As for the ancilla qubit $C_{N+1}$, it keeps state $|0\rangle$ after the addition of two input $N$-bit binary integers.

We present in Alg.(\ref{alg_adder}) the algorithm to implement quantum adders supporting multiple bits, where the notation $\gets$ statements that the qubits (on the left hand) are converted or mapped to the quantum state (on the right hand), or the value (on the right hand) is stored in the variable (on the left hand). 

\begin{algorithm}[H]
\caption{Quantum Adders for Multiple Bits}
\label{alg_adder}
\begin{algorithmic}
\State {\bf Input:} Two N-bit binary integers: 
\State \qquad $(a_{N-1}\cdots a_0)_{bin}$, $(b_{N-1}\cdots a_0)_{bin}$.
\State  $A_{N-1},\cdots,A_0\gets |a_{N-1}\cdots a_0\rangle$,
\State  $B_{N-1},\cdots,B_0,B_{-1}\gets |b_{N-1}\cdots b_00\rangle$,
\State  $C_{N},\cdots,C_{0}\gets |0\cdots 0\rangle$.
\State $n \gets 0$
\While{$n < N$}
    \State {\bf do} $\hat{P}_{I}$ or $\hat{P}_{II}$ on 
    \State \qquad $A_n, B_n, C_n, C_{n+1}, C_{n+2}$, 
    \State \qquad $C_n\gets |s_n\rangle$, $C_{n+1}\gets |c^{out}_{n}\rangle$.
    \State {\bf do} $n \gets n+1$
\EndWhile
\State {\bf Output:} $C_N,\cdots,C_0\gets |s_N\cdots s_0\rangle$.
\end{algorithmic}
\end{algorithm}

We denote the quantum adder in Alg.(\ref{alg_adder}) as $P_I(A_{N-1},\cdots,A_0;C_{N},\cdots,C_0;B_{N-1},\cdots,B_0)$ and $P_{II}(A_{N-1},\cdots,A_0;B_{N-1},\cdots,B_0;C_{N},\cdots,C_0)$, where the subscripts refer to the first type or second type quantum adder as shwon in Eq.(\ref{eq_p1_I},\ref{eq_p1_II}).
For simplicity, the involved qubits, and the corresponding quantum states, are given from left columns to right columns, from top to bottom, as depicted in Fig.(\ref{fig_adder_qubit}c,d).
We have
\begin{equation}
    \begin{split}
        P_I|a_{N-1}\cdots a_0,0\cdots 0,b_{N-1}\cdots b_0\rangle
    \\
    =
    |a_{N-1}\cdots a_0, s_N\cdots s_0, b_{N-1}\cdots b_0\rangle
    \end{split}
\end{equation}
and
\begin{equation}
    \begin{split}
        P_{II}|a_{N-1}\cdots a_0,b_{N-1}\cdots b_0, 0\cdots 0\rangle
    \\
    =
    |a_{N-1}\cdots a_0, b_{N-1}\cdots b_0, s_N\cdots s_0\rangle
    \end{split}
\end{equation}
where for simplicity, the involved qubits are omitted, and the notation of quantum adders are abbreviated as $P_{I}$, $P_{II}$.

To add the two $N$-bit binary integers, the quantum one-bit full adder is applied $N$ times, and the full quantum circuit can be decomposed into $\mathcal{O}(N)$ one-qubit or CNOT gates.
The proposed quantum adder supporting multi-bit addition works like the typical ripple-carry adder, where each carry bit "ripples" to the next full adder.
Thus, we have to repeat the quantum one-bit full adders step by step, and the time complexity is also $\mathcal{O}(N)$, and the circuit depth is of the same order.

\subsection{Special Quantum Adders Supporting Multiple Bits}
In Sec.(\ref{Quantum Adders Supporting Multiple Bits}) we present two types of quantum adders that support multiple bits.
These two quantum adders are both feasible implementations of the binary addition as described in Eq.(\ref{eq_adder_s},\ref{eq_adder_cout}).
Even though, these quantum adders can be `expensive' in more intricate arithmetic logic circuits, especially for iterative additions. 
Recalling Fig.(\ref{fig_adder_qubit}c,d), the states of qubits $A$, $B$ that correspond to the input integers do not change after applying the full circuit, and the outputs correspond to the final state of qubits $C$, which are initialized at state $|0\rangle$ at the beginning. 
Thus, every time we add two $N$-bit binary integers, $N+1$ qubits are required to store the output.
In intricate arithmetic logic circuits, such as multipliers and dividers as discussed in Sec.(\ref{sec_multiplier}) and Sec.(\ref{sec_divider}), there are often iterative additions, and it is hardly affordable to assign additional qubits to store the temporary outputs of the iterative additions.

\begin{figure*}
    \centering
    \includegraphics[width=0.85\textwidth]{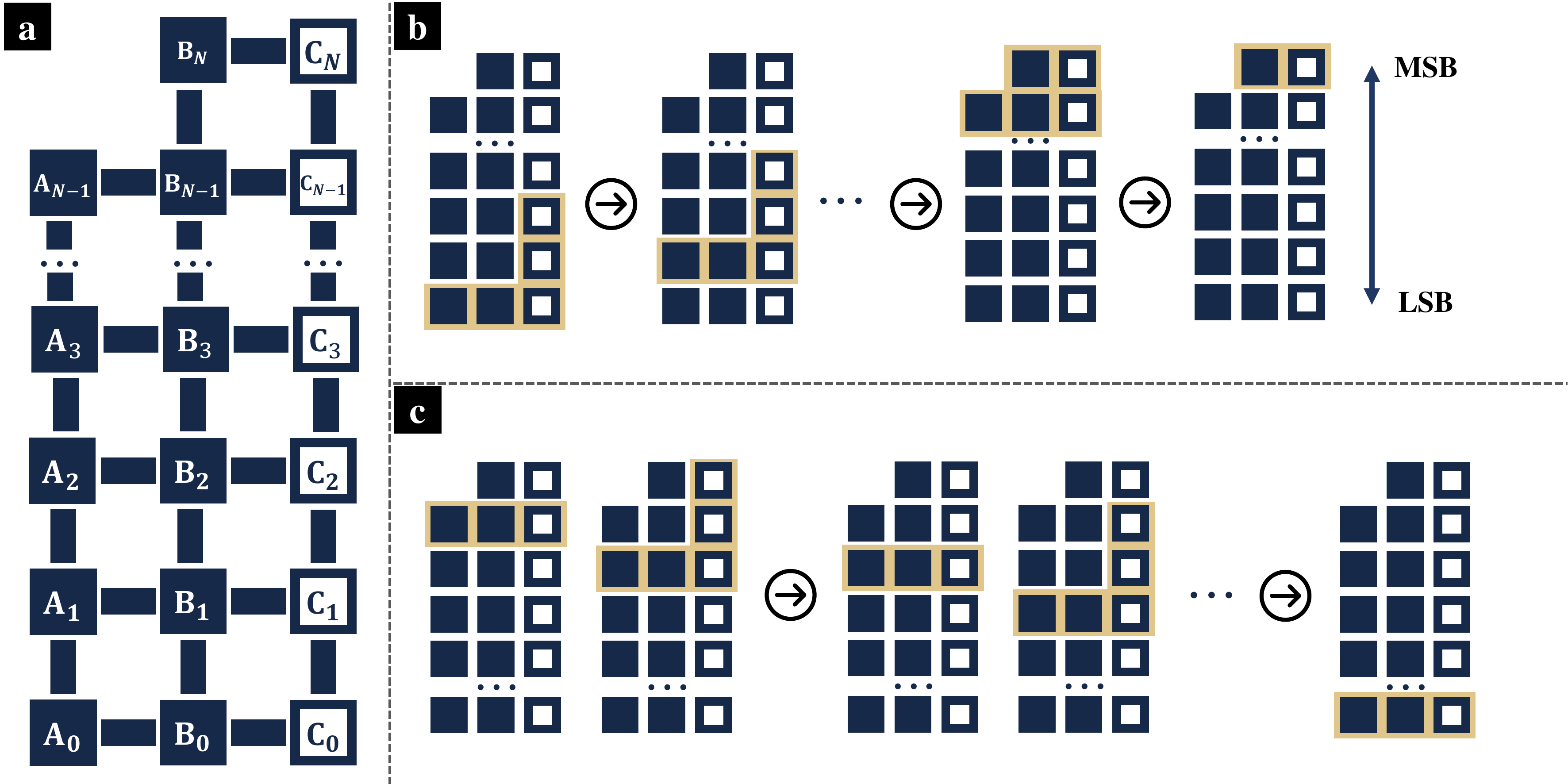}
    \caption{
    (a) Required qubits and their connectivity for the special quantum adder, $P_{III}$.
    Squares in the graph represent physical qubits and edges are the allowed interactions.
    (b,c) Diagram of the activated qubits in $P_{III}$.
    The squares correspond to the qubits as depicted in (a), and the activated qubits are highlighted in light yellow.
    }
    \label{fig_qadder_sp}
\end{figure*}

To address this issue, in this section we propose a special quantum adder, where no additional qubits are required to store the output.
For simplicity, the special quantum adder is denoted as $P_{III}$, referring to the third type quantum adder proposed in this paper.
In Fig.(\ref{fig_qadder_sp}a) we present the required qubits and their connectivity for $P_{III}$.
The inputs are two $N$ bit integers $(a_{N-1}\cdots a_0)_{bin}$, $(b_{N-1}\cdots b_0)_{bin}$, mapped to $2N$ qubits denoted as $A_{N-1},\cdots,A_0$ and $B_{N-1}, \cdots, B_0$.
We need one more qubit $B_N$ to store the outputs, as the final output is a $N+1$-bit binary integer.
Qubits $B_N, \cdots,B_0$ will be converted to state $|s_N,\cdots,s_0\rangle$ at end, indicating the output sum $(s_{N}\cdots s_0)_{bin}$.
Additionally, there are $N+1$ ancilla qubits denoted as $C_{N-1},\cdots,C_0$.
These qubits are initialized at $|0\rangle$, and will still be at state $|0\rangle$ by the end.
Thus, we can still use these ancilla qubits in the succeeding addition operations.

\begin{figure*}[th]
    \centering
    \includegraphics[width=0.9\textwidth]{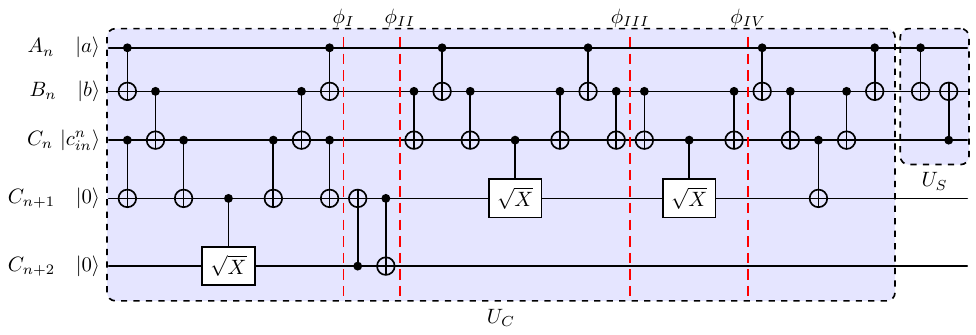}
    \caption{Detailed quantum circuits of operation $U_C$ and $U_S$.
     The subscripts indicate the output carry (C) and the output sum (S).
    $U_C$ computes the output carry, and the operations before $\phi_{IV}$ is exactly the same to $P_{II}$, as depicted in Fig.(\ref{fig_qadder_2}).
    $U_S$ computes the output sum, formed by two CNOT gates.}
    \label{fig_qadder_sp1}
\end{figure*}

There are two main tasks in the implementation of $P_{III}$.
One task is to calculate the output carry and the output sum.
The other is to reset the ancilla qubits to state $|0\rangle$.
As depicted in Fig.(\ref{fig_qadder_sp1}), we present the key operations in $P_{III}$, denoted as $U_C$ and $U_S$, where the subscripts indicate the output carry (C) and the output sum (S).
$U_C$ is designed to work as Eq.(\ref{eq_qadder_c}).
The operations before $\phi_{IV}$ is exactly the same to $P_{II}$, as depicted in Fig.(\ref{fig_qadder_2}).
After $\phi_{IV}$, the first three CNOT gates realize operation $X^{a\oplus b}$ on qubit $C_{n+1}$, and the other two CNOT gates are included to reset qubit $A_n$ and $B_n$ to state $|a\rangle$ and $|b\rangle$.
Therefore we have
\begin{equation}
    \begin{split}
        &U_C(A_n,B_n,C_{n+2},C_{n+1},C_{n})
    |a_n,b_n,0,0,c^n_{in}\rangle
    \\
    =&
    |a_n,b_n,0,c^{n+1}_{in},c^n_{in}\rangle
    \end{split}
\end{equation}
where the output carry of the $n^{th}$ bit $c^n_{out}$ is also the input carry of the $(n+1)^{th}$ bit $c^{n+1}_{in}$.
Notice that $U_C$ is reversible, we can reset qubit $C_{n+1}$ to $|0\rangle$ by applying $U_C^\dagger$, as
\begin{equation}
    \begin{split}
        &U_C^\dagger(A_n,B_n,C_{n+2},C_{n+1},C_{n})
        |a_n,b_n,0,c^{n+1}_{in},c^n_{in}\rangle
        \\
        =&|a_n,b_n,0,0,c^n_{in}\rangle
    \end{split}
\end{equation}
As for $U_S$, it works as Eq.(\ref{eq_qadder_s}), and we have
\begin{equation}
    U_S(A_n,B_n,C_n)|a_n,b_n,c_{in}^n\rangle
    =|a_n,s_n,c_{in}^n\rangle
\end{equation}

\begin{algorithm}[H]
\caption{Quantum Adder $P_{III}$}
\label{alg_adder_III}
\begin{algorithmic}
\State {\bf Input:} Two N-bit binary integers: 
\State \qquad $(a_{N-1}\cdots a_0)_{bin}$, $(b_{N-1}\cdots a_0)_{bin}$.
\State  $A_{N-1},\cdots,A_0\gets |a_{N-1}\cdots a_0\rangle$,
\State  $B_N,B_{N-1},\cdots,B_0\gets |0b_{N-1}\cdots b_0\rangle$,
\State  $C_{N},\cdots,C_{0}\gets |0\cdots 0\rangle$.
\State $n \gets 0$
\While{$n < N$}
    \State {\bf do} $U_C$ on 
    \State \qquad $A_n, B_n, C_n, C_{n+1}, C_{n+2}$, 
    \State \qquad $C_n\gets |s_n\rangle$, $C_{n+1}\gets |c^{out}_{n}\rangle$.
    \State {\bf do} $n \gets n+1$
    \EndWhile
\State {\bf do} SWAP on $B_N$, $C_N$,
\State $B_N\gets |s_N\rangle$, $C_N\gets |0\rangle$.
\State $n \gets N-1$
\While{$n > 0$}
    \State {\bf do} $U_S$ on 
    \State \qquad $A_n, B_n, C_n$, 
    \State \qquad $B_n\gets |s_n\rangle$, $C_{n}\gets |0\rangle$.
    \State {\bf do} $U_C^\dagger$ on 
    \State \qquad $A_{n-1}, B_{n-1}, C_{n+1}, C_{n}, C_{n-1}$,
    \State \qquad $C_n\gets |0\rangle$.
    \State {\bf do} $n \gets n-1$
\EndWhile
\State {\bf do} CNOT on 
    \State \qquad $A_0, B_0$ ($A_0$ is control qubit), 
    \State \qquad $B_0\gets |s_0\rangle$.
\State {\bf Output:} $B_N,\cdots,B_0\gets |s_N\cdots s_0\rangle$.
\end{algorithmic}
\end{algorithm}

We present in Alg.(\ref{alg_adder}) the algorithm to implement quantum adder $P_{III}$, which works as follows.
The first step is to apply $U_C$ for each digit, from LSB to MSB.
The $U_C$ operations prepare qubit $C_n$ at state $|c^{n}_{in}\rangle$ ($n=0,1,\cdots,N$), while qubits $A_n$ and $B_n$ are kept as the input state $|a_n\rangle$ and $|b_n\rangle$ ($n=0,1,\cdots,N-1$), and $B_N$ is still at state $|0\rangle$.
Then apply a SWAP gate between qubits $C_N$ and $B_N$, swapping the quantum states of these two qubits.
Thus, $B_N$ is converted to state $|c^{N}_{in}\rangle$, and $C_N$ is reset to state $|0\rangle$.
Notice that $|c^{N}_{in}\rangle$, or $|c^{N-1}_{out}\rangle$, is also the MSB of output sum, $|s_{N}\rangle$.
By this mean, we have qubits $C_N$ and $B_N$ at the correct output states.
In Fig.(\ref{fig_qadder_sp}b) we present the activated qubits till this step.
The activated qubits are marked with the light yellow background, where each square represents a physical qubit, corresponding to Fig.(\ref{fig_qadder_sp}a).
Hereafter, our aim is to convert qubits $B_n$ and $C_n$ ($n=0,1,\cdots,N-1$) to the correct output states.
For each digit ($n=N-1, N-2,\cdots, 1$, from top to bottom), apply $U_S$ on qubits $A_n$, $B_n$, and $C_n$, and then apply $U_C^{\dagger}$ to qubits $A_{n-1}$, $B_{n-1}$, $C_{n+1}$, $C_{n}$ and $C_{n-1}$.
$U_S$ converts qubit $B_n$ to the correct output $|s_n\rangle$, while $U_C^{\dagger}$ reset qubit $C_n$ to state $|0\rangle$.
By the end, for qubits $A_0$, $B_0$, $C_0$ that represent the LSB, we do not need to apply the full $U_S$.
Instead, a single CNOT gate on qubits $A_0$, $B_0$ ($A_0$ is control qubit) can convert $B_0$ to the correct output $|s_0\rangle$, as the input carry for the least significant bit is 0.
The activated qubits are marked with the light yellow background in Fig.(\ref{fig_qadder_sp}c).
In brief, $P_{III}(A_{N-1}, \cdots, A_0; B_{N}, \cdots, B_0; C_{N}, \cdots, C_0)$ works as
\begin{equation}
    \begin{split}
        P_{III}|a_{N-1}\cdots a_0,0b_{N-1}\cdots b_0,0\cdots 0\rangle
        \\
        =|a_{N-1}\cdots a_0,s_Ns_{N-1}\cdots s_0, 0\cdots 0\rangle
    \end{split}
\end{equation}
where the involved qubits of $P_{III}$ are omitted for simplicity.

There are 23 $CNOT$ gates and 3 $C\sqrt{X}$ gates in the decomposition of $U_C$, and 2 $CNOT$ gates in $U_S$.
In the quantum adder $P_{III}$ that adds two $N$-bit integers, $U_C$ is applied $N$ times, $U_C^\dagger$ is also applied $N$ times, and $U_S$ is applied $N-1$ times.
Moreover, there is a SWAP gate between qubits $B_N$ and $C_N$ (can be decomposed as 2 $CNOT$ gates, as one qubit is at $|0\rangle$ initially), and one single $CNOT$ gate between $A_0$ and $C_0$.
In total, $P_{III}$ of two $N$-bit integers can be decomposed into $\mathcal{O}(N)$ two-qubit gates.
As these operations can not run in parallel, the time complexity is also of the order $\mathcal{O}(N)$.

Notice that in $P_{III}$, the two given terms are not equivalent.
The special quantum adder does not change one input denoted as $(a_{N-1}\cdots a_0)_{bin}$.
On the contrary, $(b_{N-1}\cdots b_0)_{bin}$ is replaced by the output sum.
To calculate a succession of additions, it is convenient to map the new terms to column $A$, and use column $B$ to store the temporary results.

\section{Two's Complement and Quantum Subtractor}
\label{sec_subtractor_main}
Two's complement is the most common method of representing signed (positive, negative, and zero) integers in classical computers or processors\cite{burks1946preliminary, horowitz1989art}.
In two's complement representation, the value $\omega$ of a $N+1$ bit integer $(a_{N},a_{N-1}\cdots a_0)_{bin}$ is given as 
\begin{equation}
    \omega = -2^Na_N+\sum_{j=0}^{N-1}2^ja_j
    \label{eq_twos_omega}
\end{equation}
where $a_N$ is sign bit indicating the sign, $a_N=0$ for non-negative and $a_N=1$ for negative integer.
The two's complement representation of a non-negative number is just its ordinary binary representation, with sign bit 0.
The two's complement representation of a negative binary number can be obtained by two steps: Firstly all bits are inverted, or "flipped", and the value of 1 is then added to the resulting value, ignoring the overflow (which occurs when taking the two's complement of 0).
Recently, a variety of quantum approaches have been proposed to compute the two's complement representation on quantum computers\cite{chaudhuri2017novel, orts2019optimized}.
Even though, it is still challenging to implement these circuits on near-term quantum computers, mainly due to the required two- or multi-qubit gates among remote qubits.
In this section, we present a feasible quantum circuits that calculates the two's complement of a given signed integer.
Moreover, we then demonstrate how to use the method of two's complement to implement subtraction on quantum adders.
Similar to the quantum adders as discussed in Sec.(\ref{sec_adder}), the proposed circuits utilize only Pauli-X gates, CNOT gates, and $C\sqrt{X}$ (CSX) gates, and all two-qubit gates are operated between nearest neighbor qubits. 
Therefore, it is feasible to implement these operations on near-term quantum computers, where qubits are arranged in two-dimensional arrays.

\subsection{Obtain the Two's Complement of A Negative Integer}
\label{sec_twos}
Hereafter we present the quantum operation that generates the two's complement representation of a given negative integer.
The input is a $N+1$-bit binary string, $a_Na_{N-1}\cdots a_0$, where the sign bit $a_N=1$ indicates that it is negative, and $(a_{N-1}\cdots a_0)_{bin}$ is the absolute binary representation of the given integer.
The inputs are mapped to $N+1$ qubits $A_N,\cdots,A_0$ as $|a_Na_{N-1}\cdots a_0\rangle$.
Our aim is to convert the inputs into the correct output state $|\Tilde{a}_N\Tilde{a}_{N-1}\cdots \Tilde{a}_0\rangle$, where $(\Tilde{a}_N, \Tilde{a}_{N-1}\cdots \Tilde{a}_0)$ is the two's complement representation, the sign bit is still $\Tilde{a}_N=1$.
Recalling Eq.(\ref{eq_twos_omega}), we have
\begin{equation}
    -2^N+\sum_{j=0}^{N-1}2^j\Tilde{a}_j
    = -\sum_{j=0}^{N-1}2^ja_j
    \label{eq_twos_q}
\end{equation}

Similar to the procedure on classical computers, there are mainly two steps to obtain $|\Tilde{a}_N\Tilde{a}_{N-1}\cdots \Tilde{a}_0\rangle$.
The first step is to invert all bits of the absolute, $a_{N-1}\cdots a_0$.
The second step is to add 1 to the entire inverted number, ignoring any overflow.
The first step can be implemented by $N$ Pauli-X gates acting on qubits $A_{N-1},\cdots,A_0$, which convert these qubits to state $|\Bar{a}_{N-1}\cdots \Bar{a}_0\rangle$, where
\begin{equation}
    |\Bar{a}_n\rangle=X|a_n\rangle, \qquad n=0,1,\cdots N-1
\end{equation}
The second step, theoretically, can be implemented by $P_{II}$ or $P_{III}$, where $1$ is treated as one term in addition.
Notice that in the $N$-bit representation of 1, all bits except the LSB is 0, which enables us to implement the second step using a shallower circuit instead of directly applying $P_{II}$ or $P_{III}$.
For simplicity, we denote the quantum operation that calculates the sum of $(\Bar{a}_{N-1}\cdots \Bar{a}_0)_{bin}$ and 1 as $P_{+1}$, where the input is a $N$-bit binary integer denoted as $(\Bar{a}_{N-1}\cdots \Bar{a}_0)_{bin}$.

Recalling the implementation of $P_{II}$ or $P_{III}$ in Sec.(\ref{sec_adder}), there are 3 inputs for each digit except the LSB, the input carry and two bits from the input integers.
In $P_{+1}$, however, there are always 2 inputs for each digit.
For LSB, the inputs are 1 and $b_0$.
For the $n^{th}$ digits ($n=1,2,\cdots N-1$), the inputs are the input carry $\Tilde{c}^n_{in}$ and $\Bar{a}_n$.
Thus, we do not need to apply the quantum one-bit full adder.
For each digit, $P_{+1}$ acts the addition as
\begin{equation}
    \Tilde{a}_n = \Tilde{c}_{in}\oplus \Bar{a}_n
    \label{eq_twos_an}
\end{equation}
\begin{equation}
    \Tilde{c}^n_{out} 
    =\Tilde{c}^n_{in}\cdot \Bar{a}_n
    \label{eq_twos_cn}
\end{equation}
where the sum is denoted as $\Tilde{a}$, and the carry is denoted as $\Tilde{c}$ to avoid confusions with Eq.(\ref{eq_adder_s},\ref{eq_adder_cout}).
The output sum $(\Tilde{a}_N\cdots\Tilde{a}_0)_{bin}$ is a $N+1$-bit integer, and 
\begin{equation}
    (\Tilde{a}_N\cdots\Tilde{a}_0)_{bin}
    =
    (\Bar{a}_{N-1}\cdots\Bar{a}_0)_{bin}+1
    \label{eq_plus1}
\end{equation}
where the MSB $\Tilde{a}_N$ represents the overflow of the addition.
For clarity, here we study the simpler case, including the overflow.
Later will give the operation that ignoring the overflow, which is integral to obtain the two's complement representation.

Recalling the implementation of $P_{III}$, $P_{+1}$ can be treated as a special formation of $P_{III}$, where one input integer is always 1.
Similarly, we can rewrite the addition of $P_{+1}$ in the context of quantum,
\begin{equation}
    |\Tilde{a}_n\rangle
    =
    X^{\Tilde{c}^n_{in}}|\Bar{a}_n\rangle
    \label{eq_s_plus1}
\end{equation}
\begin{equation}
    |\Tilde{c}^n_{out}\rangle
    =
    X^{\frac{3}{2}(\Bar{a}_n\oplus \Tilde{c}^n_{in})+\frac{1}{2}(\Bar{a}_n+ \Tilde{c}^n_{in})}
    \label{eq_c_plus1}
\end{equation}

\begin{figure*}[th]
    \centering
    \includegraphics[width=0.7\textwidth]{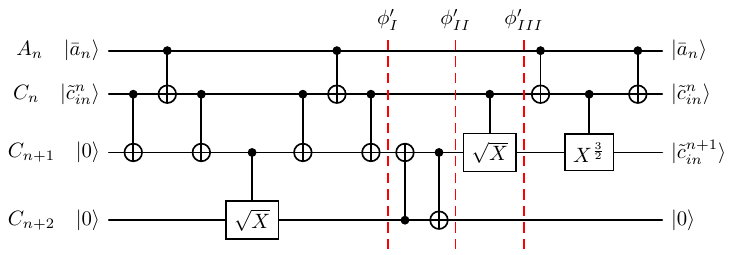}
    \caption{{\bf Detailed quantum circuits of operation $\Tilde{U}_C$.}
    Four qubits are involved in $\Tilde{U}_C$.
    Qubit $A_n$ is initialized at state $|\Bar{a}_n\rangle$, corresponding to the input, whereas qubit $C_n$ is initialized at state $|\Tilde{c}_{in}^n\rangle$, representing the input carry.
    States of qubits $A_n$, $C_n$ do not change after applying $\Tilde{U}_C$.
    Qubit $C_{n+1}$ is initialized at state $|0\rangle$, and is converted to state $|\Tilde{c}^{n+1}_{in}\rangle$ (or equivalently, $|\Tilde{c}^n_{out}\rangle$) by the end, corresponding to the output carry.
    Qubit $C_{n+2}$ is ancilla qubit, which is initialized at $|0\rangle$, and is reset at $|0\rangle$ after applying $\Tilde{U}_C$.
    The operation $CX^\frac{3}{2}$ can be implemented by a CNOT gate along with a $C\sqrt{X}$ gate.
    }
    \label{fig_qadder_plus1}
\end{figure*}

\begin{figure*}
    \centering
    \includegraphics[width=0.85\textwidth]{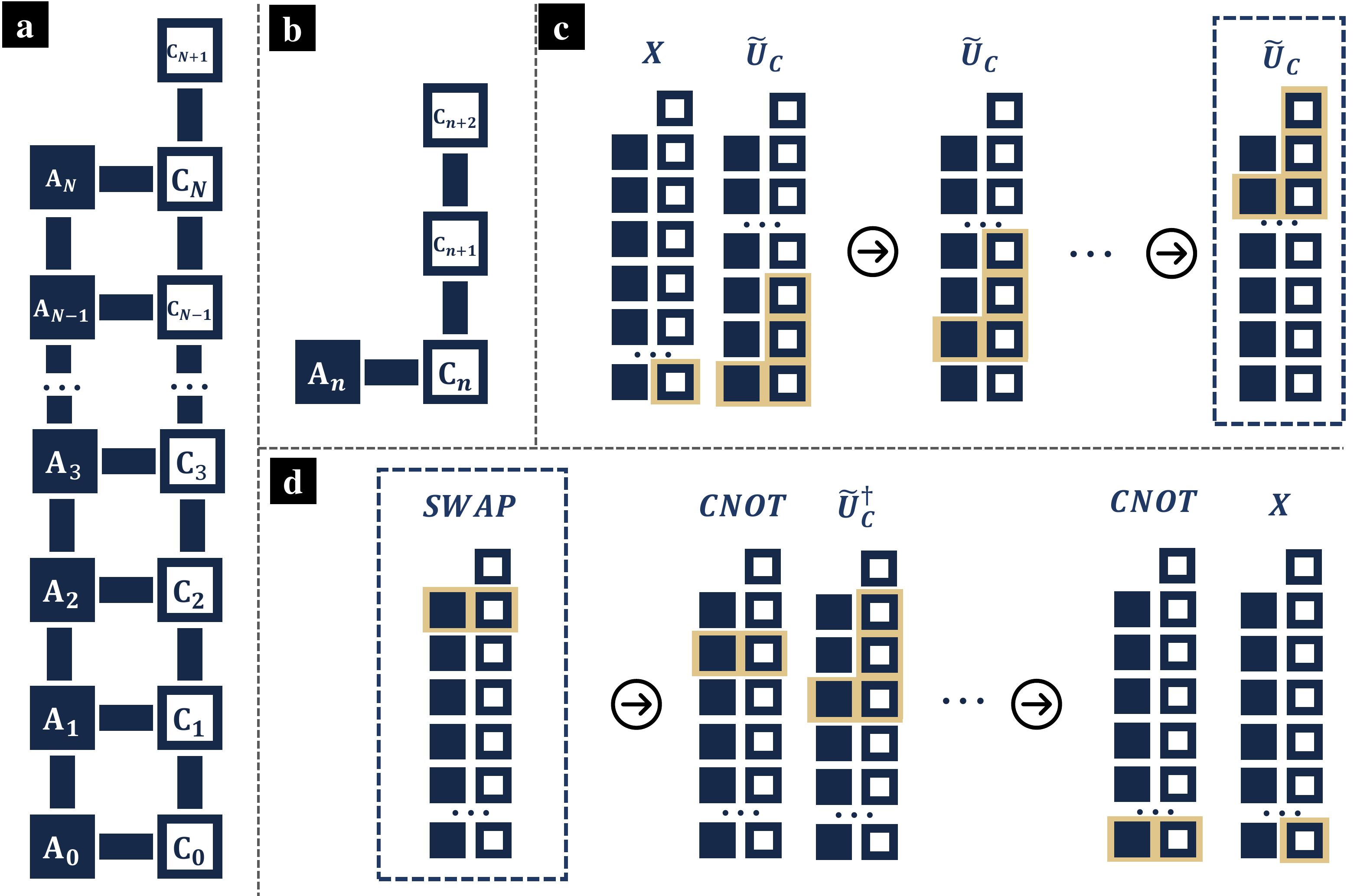}
    \caption{
    (a) Required qubits and their connectivity for the special quantum adder, $P_{+1}$ ($\Tilde{P}_{+1}$).
    Squares in the graph represent physical qubits and edges are the allowed interactions.
    (b) Required qubits and their connectivity for operation $\Tilde{U}_C$.
    (c,d) Diagram of the activated qubits in $P_{+1}$.
    The squares correspond to the qubits as depicted in (a), and the activated qubits are highlighted in light yellow.
    }
    \label{fig_plus1}
\end{figure*}

These operations can be implemented by a chain of special quantum `one-bit' adders, which works like classical half adders.
Consider 4 qubits $A_n$, $C_n$, $C_{n+1}$, $C_{n+2}$ as depicted in Fig.(\ref{fig_plus1}b).
$A_n$ is initialized at state $|\Bar{a}_n\rangle$, $C_n$ is initialized at $|\Tilde{c}^n_{in}\rangle$, $C_{n+1}$, $C_{n+2}$ are initialized at state $|0\rangle$.
Eq.(\ref{eq_s_plus1}) can be implemented by a CNOT gate.
To implement Eq.(\ref{eq_c_plus1}), we develop an operation $\Tilde{U}_C$ as depicted in Fig.(\ref{fig_qadder_plus1}).
The first part of $\Tilde{U}_C$ corresponds to operation $X^{\frac{1}{2}\Tilde{a}_n}$.
As depicted in Fig.(\ref{fig_qadder_plus1}), the operations before the dashed line $\phi'_I$ converts qubit $C_{n+2}$ to state $X^{\frac{1}{2}\Tilde{a_n}}|0\rangle$, and then next two CNOT gates are applied between $\phi'_I$ and $\phi'_{II}$, swapping the states of qubit $C_{n+1}$ and qubit $C_{n+2}$.
The succeeding $C\sqrt{X}$ gate corresponds to operation $X^{\frac{1}{2}\Tilde{c}_{in}^n}$.
By the end, the operations after $\phi'_{III}$ correspond to operation $X^{\frac{3}{2}(\Bar{a}_n\oplus \Tilde{c}^n_{in})}$.
Notice that CNOT gates and $C\sqrt{X}$ gates commute, the operation $CX^\frac{3}{2}$ can be implemented by a CNOT gate along with a $C\sqrt{X}$ gate.
$\Tilde{U}_C$ works as follows,
\begin{equation}
    \begin{split}
        &\Tilde{U}_C(A_n,C_n,C_{n+1},C_{n+2})
    |\Bar{a}_n, \Tilde{c}^n_{in}, 0, 0\rangle
    \\
    =&
    |\Bar{a}_n, \Tilde{a}_n, \Tilde{c}^{n+1}_{in}, 0\rangle
    \end{split}
\end{equation}
where $\Tilde{a}_n$ and $\Tilde{c}_n$ are given in Eq.(\ref{eq_twos_an}) and Eq.(\ref{eq_twos_cn}).

Hereafter we demonstrate the implementation of operation $P_{+1}$, which is constituted by a cascade of $\Tilde{U}_C$ and CNOT gates.
Initially, $N$ qubits $A_n$ ($n=0,1,\cdots,N-1$) is prepared at state $|\Bar{a}_n\rangle$, corresponding to the input $N$-bit integer $(0a_{N-1}\cdots a_0)_{bin}$.
There is one additional qubit $A_N$, which is prepared at $|0\rangle$, and will store the overflow of the final output.
The $N+2$ ancilla qubits $C_n$ ($n=0,1,\cdots,N+1$) are all initialized at $|0\rangle$.
Firstly, Pauli-X gate is applied on qubit $C_0$, generating the one-bit integer 1 in the addition.
Then apply $\Tilde{U}_C$ from LSB (corresponding to qubits $A_0, C_{2}, C_{1}, C_0$) to MSB (qubits $A_{N-1}, C_{N+1}, C_{N}, C_{N-1}$).
By this mean, we have all qubits $C_n$ at state $|\Tilde{c}^n_{in}\rangle$ ($n=0,\cdots,N$), corresponding to the input carry $\Tilde{c}^n_{in}$.
Next, apply a SWAP gate between $C_N$ and $A_N$.
Thus, we have qubit $A_N$ at state $|\Tilde{c}^N_{in}\rangle$, which is also the MSB of the output sum, and can be rewritten as $|\Tilde{a}_N\rangle$.
Meanwhile, ancilla qubit $C_N$ is reset to $|0\rangle$.
Iteratively, apply CNOT gates (qubits $C$ are control qubits) and $\Tilde{U}_C^\dagger$ from MSB to the second least significant bit.
By this mean, we have prepared qubit $A_n$ at state $|\Tilde{a}_n\rangle$, and reset ancilla qubit $C_n$ at state $|0\rangle$ ($n=N-1,\cdots,1$).
As for the LSB, a CNOT gate is applied on $C_0$ and $A_0$, where $C_0$ is the control qubit.
The CNOT gate converts $A_0$ to state $|\Tilde{a}_0\rangle$, corresponding to the LSB of output sum.
By the end, a Pauli-X gate is applied on qubit $C_0$, resetting it to state $|0\rangle$.
In Fig.(\ref{fig_plus1}c,d), the activated qubits in the above steps are highlighted with light yellow background, and the applied operations are marked at top.

We present in Alg.(\ref{alg_plus1}) the algorithm to implement quantum operation $P_{+1}$.
In brief, $P_{+1}(A_{N},\cdots,A_0;C_{N+1},\cdots,C_0)$ works as
\begin{equation}
    \begin{split}
        &P_{+1}|0\Bar{a}_{N-1}\cdots \Bar{a}_0,0\cdots 0\rangle
        \\
        =&|\Tilde{a}_{N}\cdots \Tilde{a}_0, 0\cdots 0\rangle
    \end{split}
\end{equation}
where the involved qubits of $P_{+1}$ are omitted for simplicity. 

\begin{figure}[ht]
\begin{algorithm}[H]
\caption{Operation $P_{+1}$, overflow included}
\label{alg_plus1}
\begin{algorithmic}
\State {\bf Input:} One N-bit binary integers: 
\State \qquad $(a_{N-1}\cdots a_0)_{bin}$.
\State $A_{N},\cdots,A_0\gets |0a_{N-1}\cdots a_0\rangle$,
\State $C_{N+1},\cdots,C_0\gets |0\cdots 0\rangle$.
\State {\bf do} Pauli-X gate on $C_0$,
\State \qquad $C_0 \gets |1\rangle$.
\State $n \gets 0$.
\While{$n < N$}
    \State {\bf do} $\Tilde{U}_C$ on 
    \State \qquad $A_n, C_{n+2}, C_{n+1}, C_n$, 
    \State \qquad $C_{n+1}\gets |\Tilde{c}_{in}^{n+1}\rangle$.
    \State {\bf do} $n \gets n+1$
    \EndWhile
\State {\bf do} SWAP gate on $A_N$, $C_N$,
\State $A_N\gets |\Tilde{a}_N\rangle$, $C_N\gets |0\rangle$.
\State $n \gets N-1$
\While{$n > 0$}
    \State {\bf do} CNOT gate on 
    \State \qquad $A_n, C_n$ ($C_n$ is control qubit), 
    \State \qquad $A_n\gets |\Tilde{a}_n\rangle$.
    \State {\bf do} $\Tilde{U}_C^\dagger$ on 
    \State \qquad $A_{n-1}, C_{n+1}, C_{n}, C_{n-1}$,
    \State \qquad $C_n\gets |0\rangle$.
    \State {\bf do} $n \gets n-1$
\EndWhile
\State {\bf do} CNOT on 
    \State \qquad $A_0, C_0$ ($C_0$ is control qubit), 
    \State \qquad $A_0\gets |\Tilde{a}_0\rangle$.
    \State {\bf do} Pauli-X gate on $C_0$,
\State \qquad $C_0 \gets |0\rangle$.
\State {\bf Output:} $A_N,\cdots,A_0\gets |\Tilde{a}_N\cdots \Tilde{a}_0\rangle$.
\end{algorithmic}
\end{algorithm}
\end{figure}

Notice that $P_{+1}$ takes the account of the overflow.
However, to obtain the two's complement of a negative integer, we need to add 1 to the inverted integer, ignoring any overflow.
In Alg.(\ref{alg_plus1_no}) we present the algorithm to implement quantum operation $\Tilde{P}_{+1}$, where the overflow is ignored.
In $\Tilde{P}_{+1}$, the operations generating $\Tilde{a}_N$ are deleted, comparing to the original version $P_{+1}$.
As the overflow is ignored, we do not need to apply $\Tilde{U}_C$ on qubits $A_{N-1}, C_{N+1}, C_{N}, C_{N-1}$, nor the SWAP gate between $C_N$ and $A_N$.
In fact, qubits $A_N$ and $C_{N+1}$ are not required in $\Tilde{P}_{+1}$.
The activated qubits in each step of $\Tilde{P}_{+1}$ still correspond to the highlighted squares in Fig.(\ref{fig_plus1}c,d), except the ones in the dashed boxes (these operations can be deleted in $\Tilde{P}_{+1}$, as the overflow is ignored).
$\Tilde{P}_{+1}$ still converts the input integer to the sum, yet in the final output quantum state, the overflow digit is not included, and we have
\begin{equation}
    \begin{split}
        &\Tilde{P}_{+1}|\Bar{a}_{N-1}\cdots \Bar{a}_0,0\cdots 0\rangle
        \\
        =&|\Tilde{a}_{N-1}\cdots \Tilde{a}_0, 0\cdots 0\rangle
    \end{split}
\end{equation}
where $\Bar{a}_{N-1}\cdots \Bar{a}_0$ and $\Tilde{a}_{N-1}\cdots \Tilde{a}_0$ correspond to the input integer and output sum in Eq.(\ref{eq_plus1}).

\begin{figure}[ht]
\begin{algorithm}[H]
\caption{Operation $\Tilde{P}_{+1}$, overflow ignored}
\label{alg_plus1_no}
\begin{algorithmic}
\State {\bf Input:} One N-bit binary integer: 
\State \qquad $(a_{N-1}\cdots a_0)_{bin}$.
\State $A_{N},\cdots,A_0\gets |0a_{N-1}\cdots a_0\rangle$,
\State $C_{N+1},\cdots,C_0\gets |0\cdots 0\rangle$.
\State {\bf* do} Pauli-X gate on $C_0$,
\State \qquad $C_0 \gets |1\rangle$.
\State $n \gets 0$.
\While{$n < N-1$}
    \State {\bf do} $\Tilde{U}_C$ on 
    \State \qquad $A_n, C_{n+2}, C_{n+1}, C_n$, 
    \State \qquad $C_{n+1}\gets |\Tilde{c}_{in}^{n+1}\rangle$.
    \State {\bf do} $n \gets n+1$
    \EndWhile
\State $n \gets N-1$
\While{$n > 0$}
    \State {\bf do} CNOT gate on 
    \State \qquad $A_n, C_n$ ($C_n$ is control qubit), 
    \State \qquad $A_n\gets |\Tilde{a}_n\rangle$.
    \State {\bf do} $\Tilde{U}_C^\dagger$ on 
    \State \qquad $A_{n-1}, C_{n+1}, C_{n}, C_{n-1}$,
    \State \qquad $C_n\gets |0\rangle$.
    \State {\bf do} $n \gets n-1$
\EndWhile
\State {\bf do} CNOT on 
    \State \qquad $A_0, C_0$ ($C_0$ is control qubit), 
    \State \qquad $A_0\gets |\Tilde{a}_0\rangle$.
    \State {**\bf do} Pauli-X gate on $C_0$,
\State \qquad $C_0 \gets |0\rangle$.
\State {\bf Output:} $A_{N-1},\cdots,A_0\gets |\Tilde{a}_{N-1}\cdots \Tilde{a}_0\rangle$.
\end{algorithmic}
\end{algorithm}
\end{figure}

In the decomposition of $P_{+1}$, $\Tilde{U}_C$ is applied $N$ times.
Additionally, there is $N-1$ CNOT gates, 2 Pauli-X gates and one SWAP gate (can be decomposed into two SWAP gates, as one input qubit is at $|0\rangle$).
Totally, $P_{+1}$ can be decomposed into $\mathcal{O}(N)$ two-qubit gates, along with 2 Pauli-X gates.
These operations are applied step by step instead of in parallel.
Therefore, the overall time complexity of $P_{+1}$ is $\mathcal{O}(N)$.
Similarly, the time complexity of $\Tilde{P}_{+1}$ is also $\mathcal{O}(N)$.

Herein we can give the procedure to obtain the two's complement representation of a given negative integer.
The input is a $N$-bit integer $(a_{N-1}\cdots a_0)_{bin}$, mapped to $N$ qubits $A_{N-1},\cdots,A_0$.
There are $N+2$ ancilla qubits $C_{N+1},\cdots,C_0$.
Additionally, we need one more qubit denoted as $A_N$, which indicates the sign of the output binary integer.
The connectivity of these $2N$ qubits are as depicted in Fig.(\ref{fig_plus1}a), where $C_{N+1}$ is not necessary.
At beginning, $A_{N-1},\cdots,A_0$ are initialized at state $|a_N\cdots a_0\rangle$, whereas $A_N$, $C_{N+1},\cdots,C_0$ are all initialized at state $|0\rangle$.
The two's complement representation can be obtained in two steps.
Firstly, apply Pauli-X gate on qubits $A_{N},\cdots,A_0$.
Next, apply $\Tilde{P}_{+1}$ on qubits $A_{N},\cdots,A_0$, and $C_{N+1},\cdots,C_0$.
These two steps corresponds to the classical approach to obtain the two's complement representation.
The first step `flips' each bit, and the sign bit is also set as 1.
The second step add 1 to the inverted integer, ignoring the overflow.
At the end, we have qubits $A_{N},\cdots,A_0$ at state $|\Tilde{a}_N\Tilde{a}_{N-1}\cdots \Tilde{a}_0\rangle$, where $\Tilde{a}$ is given as Eq.(\ref{eq_twos_q}), and the MSB $\Tilde{a}_N=1$ indicates that the integer is negative.
Meanwhile, the ancilla qubits $C_{N+1},\cdots,C_0$ are still at state $|0\rangle$.
As for the time complexity, the Pauli-X gate is applied for $N$ times, and $\Tilde{P}_{+1}$ is applied once.
In total, the time complexity is $\mathcal{O}(N)$.

$\Tilde{P}_{+1}$ is a key operation in the implementation of quantum subtractors, as discussed in Sec.(\ref{sec_subtractor}).
Moreover, later in Sec.(\ref{sec_twos3}), we will show that $\Tilde{P}_{+1}$ is integral in the operation that computes the two's complement representation of arbitrary signed integers, where the initializing process is replaced (The replaced lines are noted with * and ** in Alg.(\ref{alg_plus1_no}).

\subsection{Implement Subtraction with Quantum Adder}
\label{sec_subtractor}
Classical computers usually use the method of complements to implement subtraction with adders.
Denote $(d_{N}\cdots d_0)_{bin}$ as the difference of minuend $(a_{N-1}\cdots a_0)_{bin}$ and subtrahend $(b_{N-1}\cdots b_0)_{bin}$, 
\begin{equation}
    (d_{N}\cdots d_0)_{bin}
    =
    (a_{N-1}\cdots a_0)_{bin}
    - (b_{N-1}\cdots b_0)_{bin}
    \label{eq_subtract}
\end{equation}
where $(a_{N-1}\cdots a_0)_{bin}$ and $(b_{N-1}\cdots b_0)_{bin}$ are two $N$-bit non-negative integers without sign bit.
On the contrast, $(d_{N}\cdots d_0)_{bin}$ is given in the two's complement representation.
The sign bit $d_N=1$ if the difference is negative, whereas $d_N=0$ if the difference is non-negative.
Recalling the definition of two's complement representation, we have
\begin{equation}
        -2^Nd_N+\sum_{j=0}^{N-1}2^jd_j
        =
        \sum_{j=0}^{N-1}2^ja_j
        +\sum_{j=0}^{N-1}2^j\Tilde{b}_j
        -2^N\Tilde{b}_N
    \label{eq_q_subtract}
\end{equation}
where $(\Tilde{b}_N\Tilde{b}_{N-1}\cdots \Tilde{b}_0)_{bin}$ is the two's complement representation of $(-b_{N-1}\cdots b_0)_{bin}$.
By this mean, the subtraction is converted to addition.

Similarly, it is feasible to calculate subtraction with quantum adder.
Consider $3N+3$ qubits $A_{N-1},\cdots,A_0$, $B_{N},\cdots,B_0$, $C_{N+1},\cdots,C_0$.
The connectivity is as depicted in Fig.(\ref{fig_qadder_sp}) (qubit $C_{N+1}$ connects to $C_N$, qubit $C_{N+1}$ is not included in Fig.(\ref{fig_qadder_sp})).
The inputs are two $N$-bit integers $(a_{N-1}\cdots a_0)_{bin}$ and $(b_{N-1}\cdots b_0)_{bin}$, and our aim is to obtain the difference $(d_{N}\cdots d_0)_{bin}$ as given in Eq.(\ref{eq_subtract}).
Initially, qubits $A_{N-1},\cdots,A_0$ are prepared at $|a_{N-1}\cdots a_0\rangle$, qubits $B_{N-1},\cdots,B_0$ are prepared at $|b_{N-1}\cdots b_0\rangle$, and qubits $B_N$, $C_{N},\cdots,C_0$ are prepared at $|0\rangle$.

The first step is to prepare the two's complement representation of $(-b_{N-1}\cdots b_0)_{bin}$.
As discussed in Sec.(\ref{sec_twos}), we can apply Pauli-X gates on each qubit $B$, and then apply $\Tilde{P}_{+1}$ on qubits $B_{N},\cdots,B_0$, and $C_{N+1},\cdots,C_0$.
Then we have qubits $B_{N},\cdots,B_0$ at state $|\Tilde{b}_N\Tilde{b}_{N-1}\cdots \Tilde{b}_0\rangle$, corresponds to the two's complement representation of $(-b_{N-1}\cdots b_0)_{bin}$.

Next, apply $P_{II}$ on qubits $A_{N-1},\cdots,A_0$, $B_{N-1},\cdots,B_0$, $C_{N},\cdots,C_0$.
Notice that the sign bit $B_N$ is not included in this step.
$|a_{N-1}\cdots a_0\rangle$ and $|\Tilde{b}_{N-1}\cdots \Tilde{b}_0\rangle$ are treated as two input terms in the addition, and the output is stored in qubits $C_{N},\cdots,C_0$.

Finally, apply CNOT gate on $B_N$ and $C_N$, where $B_N$ is the control qubit. By this mean, we have qubits $C_{N},\cdots,C_0$ at state $|d_{N}\cdots d_0\rangle$, where $(d_{N}\cdots d_0)_{bin}$ is the difference in two's complement representation, and $d_N$ is the sign bit.

According to Sec.(\ref{sec_twos}), the time complexity to obtain the two's complement representation is $\mathcal{O}(N)$.
The time complexity of $P_{II}$ is also $\mathcal{O}(N)$.
Thus, the total time complexity of this quantum subtractor is also $\mathcal{O}(N)$.

Sometimes we need to sum up the input signed integers iteratively, where one input is always negative and the other is not.
For example, in quantum multiplier and quantum divider as presented later in Sec.(\ref{sec_multiplier}) and Sec.(\ref{sec_divider}), we iteratively calculate the contribution for each digit, where we need to add up the temporary results (signed, can be negative) along with the temporary sums.
In this cases, it is hardly wisdom to apply the quantum subtractor based on $P_{II}$, as it is too expensive to assign qubits to store these temporary sums.
To address this issue, in Alg.(\ref{alg_adder_III_sign}), we present the quantum adder for signed integers, which is denoted as $\Tilde{P}_{III}$.
There are two signed integers as input, $(a_{N-1}, a_{N-2}\cdots a_0)_{bin}$, and $(b_{N-1}, b_{N-2}\cdots b_0)_{bin}$, where $a_{N-1}$, $b_{N-1}$ are sign bits, and the integers are already given in the two's complement representation.
The required connectivity is as depicted in Fig.(\ref{fig_qadder_sp}a), where qubit $B_N$ is not necessary.
The given signed integers are mapped to qubits $A_{N-1},\cdots,A_0$, $B_{N-1},\cdots,B_0$, where the outputs are also stored in $B_{N-1},\cdots,B_0$.
$\Tilde{P}_{III}$ works like $P_{III}$, but ignoring the overflow.
The output is also a signed integer in the two's complement representation.
The time complexity of $\Tilde{P}_{III}$ is also $\mathcal{O}(N)$.

\begin{figure}[ht]
\begin{algorithm}[H]
\caption{Adder $\Tilde{P}_{III}$ for signed integers}
\label{alg_adder_III_sign}
\begin{algorithmic}
\State {\bf Input:} Two signed N-digit binary integers (in two's complement representation): 
\State \qquad $A_{N-1},\cdots,A_0\gets |a_{N-1}, a_{N-2}\cdots a_0\rangle$,
\State \qquad $B_{N-1},\cdots,B_0\gets |b_{N-1}, b_{N-2}\cdots b_0\rangle$,
\State \qquad $C_N,\cdots,C_0\gets |0\cdots 0\rangle$.
\State $n \gets 0$
\While{$n < N-1$}
    \State {\bf do} $U_C$ on 
    \State \qquad $A_n, B_n, C_n, C_{n+1}, C_{n+2}$, 
    \State \qquad $C_n\gets |s_n\rangle$, $C_{n+1}\gets |c_{in}^{n+1}\rangle$.
    \State {\bf do} $n \gets n+1$
    \EndWhile
\State {\bf do} CNOT on $C_{N-1}$ (Control), $B_{N-1}$ (Target),
\State {\bf do} CNOT on $A_{N-1}$ (Control), $B_{N-1}$ (Target),
\State $B_{N-1}\gets |a_{N-1}\oplus b_{N-1}\oplus c^{N-1}_{in}\rangle$.
\State {\bf do} $U_C^\dagger$ on 
    \State \qquad $A_{N-2}, B_{N-2}, C_{N}, C_{N-1}, C_{N-2}$,
    \State \qquad $C_{N-1}\gets |0\rangle$.
\State $n \gets N-2$
\While{$n > 0$}
    \State {\bf do} $U_S$ on 
    \State \qquad $A_n, B_n, C_n$, 
    \State \qquad $B_n\gets |s_n\rangle$, $C_{n}\gets |0\rangle$.
    \State {\bf do} $U_C^\dagger$ on 
    \State \qquad $A_{n-1}, B_{n-1}, C_{n+1}, C_{n}, C_{n-1}$,
    \State \qquad $C_n\gets |0\rangle$.
    \State {\bf do} $n \gets n-1$
\EndWhile
\State {\bf do} CNOT on 
    \State \qquad $A_0, B_0$ ($A_0$ is control qubit), 
    \State \qquad $B_0\gets |s_0\rangle$.
\State {\bf Output:} $B_N,\cdots,B_0\gets |s_N\cdots s_0\rangle$.
\end{algorithmic}
\end{algorithm}
\end{figure}

At the end of this subsection, we will demonstrate a useful feature of the operation $\Tilde{P}_{III}$.
For an arbitrary non-negative binary integer $(a_{N-1}\cdots a_0)_{bin}$, we have
\begin{equation}
    \begin{split}
        &(0,a_{N-1}\cdots a_0)_{bin}+(1,\Tilde{a}_{N-1}\cdots \Tilde{a}_0)_{bin}
        \\
        =&(1,0,0\cdots0)_{bin}
    \end{split}
\end{equation}
where $(1,\Tilde{a}_{N-1}\cdots \Tilde{a}_0)_{bin}$ is the two's complement representation of $(-a_{N-1}\cdots a_0)_{bin}$.
In the summation, all digits are 0, except the first 1 which is an overflow.
For the quantum adder $P_{III}$, we have
\begin{equation}
    \begin{split}
        &P_{III}|0a_{N-1}\cdots a_0,01\Tilde{a}_{N-1}\cdots \Tilde{a}_0,0\cdots 0\rangle
        \\
        =&|0a_{N-1}\cdots a_0,100\cdots0, 0\cdots 0\rangle
    \end{split}
\end{equation}
where in the second input term, $01\Tilde{a}_{N-1}\cdots \Tilde{a}_0$, the first digit 0 is a `blank' to store the overflow, and the second digit 1 is the sign bit.
As for the quantum adder $\Tilde{P}_{III}$, we have
\begin{equation}
    \begin{split}
        &\Tilde{P}_{III}|0a_{N-1}\cdots a_0,1\Tilde{a}_{N-1}\cdots \Tilde{a}_0,0\cdots 0\rangle
        \\
        =&|0a_{N-1}\cdots a_0,00\cdots0, 0\cdots 0\rangle
    \end{split}
    \label{eq_p3tilde}
\end{equation}
As $\Tilde{P}_{III}$ ignores the overflow, the output is just 0.
This feature is of great convenience in the implementation of quantum multiplier and divider, as discussed later in Sec.(\ref{sec_multiplier}) and Sec.(\ref{sec_divider}).

\subsection{Obtain the Two's Complement of An Arbitrary Signed Integer}
\label{sec_twos3}

In Sec.(\ref{sec_twos}), we present how to obtain the quantum state that corresponds to the two's complement of a given negative integer.
In this subsection we focus on a more general case, where the input is an arbitrary signed integer, either negative or non-negative.

Consider $N+1$ qubits $A_{N}, A_{N-1},\cdots,A_0$.
Qubits $A_{N-1},\cdots,A_0$ are initialized at state $|a_{N-1}\cdots a_0\rangle$, corresponding to the absolute value $(a_{N-1}\cdots a_0)_{bin}$.
Qubit $A_N$ is initialized at state $|a_N\rangle$, where $a_N$ is the sign bit, $a_N=0$ if the input is non-negative, and $a_N=1$ if the input is negative.
Our aim is to design a unitary operation $U_{\pm}$, that converts qubits $A_{N},A_{N-1},\cdots,A_0$ to state $|\Tilde{a}_N\Tilde{a}_{N-1},\cdots,\Tilde{a}_0\rangle$, where $\Tilde{a}_N = a_N$ is still the sign bit.
When the input is non-negative, $a_N=0$, $U_{\pm}$ does not change anything, and we have $(\Tilde{a}_{N-1}\cdots\Tilde{a}_0)_{bin}=(a_{N-1}\cdots a_0)_{bin}$.
On the contrary, for negative input $a_N=1$, $U_{\pm}$ converts qubits $A_{N-1},\cdots,A_0$ to state $|\Tilde{a}_{N-1}\cdots \Tilde{a}_0\rangle$, where $\Tilde{a}$ is the two's complement representation of $(-a_{N-1}\cdots a_0)_{bin}$ as shown in Eq.(\ref{eq_twos_q}).

Intuitively, $U_{\pm}$ can be implemented by an intricate `$CU$' operation, where the sign bit qubit $A_N$ is the control qubit, and `$U$' is the quantum operation that generates the two's complement representation of a given negative integer, as presented in Sec.(\ref{sec_twos}).
However, due to the limitation of hardware, it is extremely challenging to implement such intricate `$CU$' operation on near-term quantum computers.

To address this issue, hereafter we present a feasible operation denoted as $U_{\pm}$.
Still, $U_{\pm}$ can be implemented with only Pauli-X gates, CNOT gates, and $C\sqrt{X}$ (CSX) gates, and all two-qubit gates are operated between nearest neighbor qubits. 
$U_{\pm}$ acts on $2N+2$ qubits.
There are $N+1$ qubits $A_N, \cdots, A_0$ representing the given integer, where $A_N$ is the sign bit.
Additionally, we need $N+1$ ancilla qubits $C_N, \cdots, C_0$.
These ancilla qubits are initialized at $|0\rangle$ at the very beginning.
The connectivity of qubits $A_N, \cdots, A_0$ and $C_N, \cdots, C_0$ are as depicted in Fig.(\ref{fig_plus1}a) (ignoring $C_{N+1}$), where each square represents a physical qubit, and the edges indicate that the neighbor qubits are connected.

Recalling the quantum operation that generates the two's complement representation of a given negative integer, there are two main steps.
Firstly, Pauli-X gates are applied on qubits $A_{N-1},\cdots,A_0$, flipping these bits.
Next, operation $\Tilde{P}_{+1}$ is applied, adding 1 to the inverted integer and ignoring the overflow.

\begin{figure}[t]
    \begin{center}
    \includegraphics[width=0.45\textwidth]{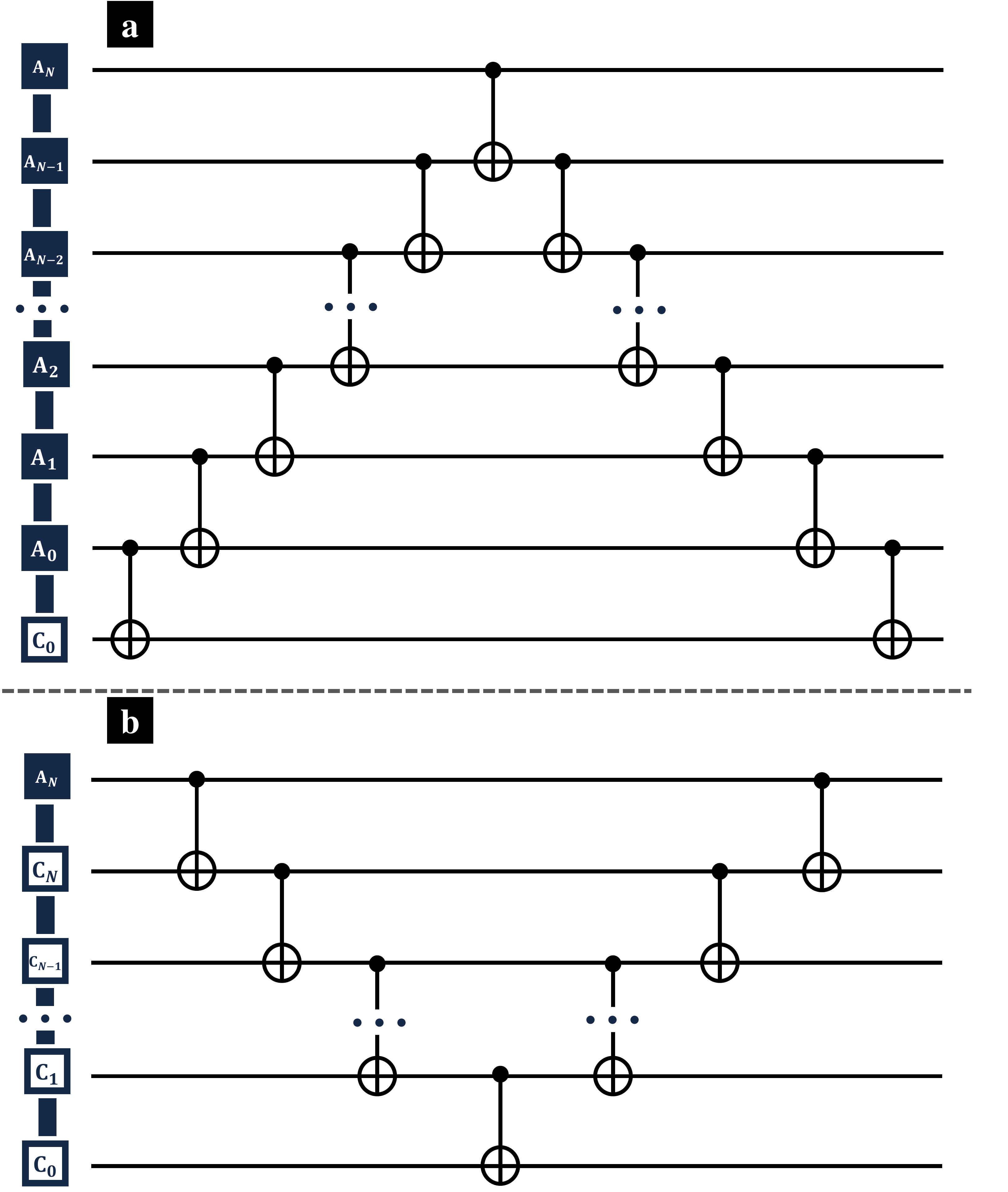}
\end{center}
    \caption{The implementation of $U_{flip}$ and $U_{res}$.}
    (a) $U_{flip}$ flips $a_{N-1},\cdots,a_0$ for negative input $a_N=1$, and meanwhile does nothing for non-negative input $a_N=0$.
    (b) $U_{res}$ resets the ancilla qubit to state $|0\rangle$.
    \label{fig_qnot}
\end{figure}

For an arbitrary signed integer, the state of sign bit $A_N$ is unknown.
In the first step, our aim is to flip these bits if the input is negative, otherwise to do nothing.
Here we apply operation $U_{flip}$ to qubits $A_N,A_{N-1},\cdots, A_0$ and ancilla qubit $C_0$.
The quantum circuit of $U_{flip}$ is as depicted in Fig.(\ref{fig_qnot}a), and the activated qubits are assigned on the left side.
Initially, qubits $A_{N},\cdots,A_0$ are prepared at input state $|a_N\cdots a_0\rangle$, while the ancilla qubit $C_0$ is at state $|0\rangle$.
There are in total $2N+1$ CNOT gates, acting on the neighbor qubits that are physically connected.
Notice that the quantum state of the control qubit does not change in a single CNOT gate.
Thus, qubit $A_N$ is always at state $|a_N\rangle$, indicating that $U_{flip}$ does not change the sign bit.
There is a CNOT gate connecting qubits $A_N$ and $A_{N-1}$, where $A_N$ is the control qubit.
By applying the CNOT gate, qubit $A_{N-1}$ is converted to state $X^{a_N}|a_{N-1}\rangle=|a_N\oplus a_{N-1}\rangle$.
Then study the output state of $A_{N-2}$.
There are two CNOT gates connecting $A_{N-1}$ and $A_{N-2}$, where qubit $A_{N-1}$ is always the control qubit.
When the first CNOT gate is applied, qubit $A_{N-1}$ is at the initial state $|a_{N-1}\rangle$, whereas when the second one is applied, $A_{N-1}$ is converted to state $|a_N\oplus a_{N-1}\rangle$.
We have
\begin{equation}
    \begin{split}
        &X^{a_N\oplus a_{N-1}}X^{a_N-1}|a_{N-2}\rangle
        \\=&|a_N\oplus a_{N-1}\oplus({a_N-1}\oplus a_{N-2})\rangle
        \\=&|a_N\oplus a_{N-2}\rangle
    \end{split}
\end{equation}
Similarly, we have qubit $A_n$, $n=0,1,\cdots,N-3$ converted to state $|a_N\oplus a_{n}\rangle$ by applying $U_{flip}$.
As for the ancilla qubit $C_0$, as it is initialized at $|0\rangle$, it is converted to output state $|a_N\rangle$.
We have
\begin{equation}
    \begin{split}
        &U_{flip}(A_N,A_{N-1},\cdots,A_0,C_0)|a_N,a_{N-1},\cdots,a_0,0\rangle
    \\=&|a_N,a_N\oplus a_{N-1},\cdots,a_N\oplus a_0,a_N\rangle
    \end{split}
    \label{eq_uflip}
\end{equation}
When the input is negative, $a_N=1$, and the output $a_N\oplus a_n$ corresponds to the flipped bit.
Otherwise, the input is non-negative, $a_N=0$, and the output is still $a_n$.
Therefore, $U_{flip}$ flips the bits $a_{N-1},\cdots,a_0$ for negative input $a_N=1$, and meanwhile does nothing for non-negative input $a_N=0$.

The second step in generating the two's complement of a negative integer is adding 1 to the inverted integer, ignoring any overflow, which can be implemented by applying the operation $\Tilde{P}_{+1}$. 
Here the given signed integer can be either negative or not, thus the original $\Tilde{P}_{+1}$ does not meet the requirement.
Recalling the operation $\Tilde{P}_{+1}$, the ancilla qubit $C_0$ is prepared at state $|0\rangle$ at beginning, then is fliped to state $|1\rangle$ (The line marked with * in Alg.(\ref{alg_plus1_no})), corresponding to the term `1' in the addition.
By the end, $C_0$ is reset to state $|0\rangle$ (The line marked with ** in Alg.(\ref{alg_plus1_no})), as it is an ancilla qubit.
For negative inputs $a_N=1$, we still need to repeat these operations.
Yet for non-negative inputs $a_N=0$, we can leave $C_0$ at the initial state $|0\rangle$, then sum up $(a_{N-1}\cdots a_0)_{bin}$ and $0$, and the output is still $(a_{N-1}\cdots a_0)_{bin}$.
As shown in Eq.(\ref{eq_uflip}), $U_{flip}$ also converts the ancilla qubit $C_0$ to state $|a_N\rangle$.
Therefore, the second step can be implemented by applying $\Tilde{P}_{+1}$  without the lines marked with * and ** in Alg.(\ref{alg_plus1_no}).
Notice that the ancilla qubit $C_0$ is still at state $|a_N\rangle$.
Thus, we need to reset $C_0$ to state $|0\rangle$.
As other ancilla qubits are all at state $|0\rangle$, we can apply $U_{res}$ as depicted in Fig.(\ref{fig_qnot}b), where there are $2N+1$ CNOT gates. 

In conclusion, the procedure of $U_{\pm}$ can be given as follows:

\noindent
{\bf 0.} Initially, a signed integer $(a_Na_{N-1}\cdots a_0)_{bin}$ is given, $a_N$ is the sign bit, for negative input $a_N=1$, otherwise $a_N=0$.
Qubits $A_N, \cdots, A_0$ are prepared at state $|a_Na_{N-1}\cdots a_0\rangle$, and ancilla qubits $C_N,\cdots,C_0$ are all prepared at state $|0\rangle$.
\\
\noindent
{\bf 1.} Firstly, apply operation $U_{flip}$ to qubits $A_N,\cdots,A_0,C_0$.
$U_{flip}$ flips $a_Na_{N-1}\cdots a_0$ if $a_N=1$, as shown in Eq.(\ref{eq_uflip}).
\\
\noindent
{\bf 2.} Next, apply $\Tilde{P}_{+1}$ as shown in Alg.(\ref{alg_plus1_no}), ignoring the lines marked with * and **.
By this mean, if $a_N=1$ we add 1 to the inverted integer, ignoring overflow.
Otherwise, we add 0 to the temporary results and nothing changes.
\\
\noindent
{\bf 3.} Finally, apply $U_{res}$ on qubits $A_N,C_N,\cdots,C_0$, resetting the ancilla qubit $C_0$ to state $|0\rangle$.

$U_{\pm}$ works as follows, 
\begin{equation}
    \begin{split}
        &U_{\pm}(A_N,\cdots,A_0;C_N,\cdots, C_0)|a_Na_{N-1}\cdots a_0, 0\cdots 0\rangle
    \\=&
    |a_N\Tilde{a}_{N-1}\cdots\Tilde{a}_0, 0\cdots 0\rangle
    \end{split}
\end{equation}
If $a_N=1$, we have $\Tilde{a}_n=\Tilde{a}_n$, where $\Tilde{a}_n$ is given in Eq.(\ref{eq_twos_q}).
Otherwise $a_N=0$, we have $\Tilde{a}_n=a_n$, which is the original input.
In conclusion, the two's complement representation $\Tilde{a}$ satisfies
\begin{equation}
    -2^Na_N+\sum_{j=0}^{N-1}2^j\Tilde{a}_j
        =
        (1-2a_N)\sum_{j=0}^{N-1}2^ja_j
    \label{eq_tilde_a}
\end{equation}
where $a_N$ is the sign bit.

As the time complexity of $U_{flip}$, $\Tilde{P}_{+1}$ and $U_{res}$ are all of the order $\mathcal{O}(N)$, we have the overall time complexity of $U_{\pm}$ is also $\mathcal{O}(N)$.
$U_{\pm}$ plays an important role the in quantum multiplier and quantum divider, which is discussed in the following sections Sec.(\ref{sec_multiplier}) and Sec.(\ref{sec_divider}).

\section{Quantum Multiplier}
\label{sec_multiplier}

Long multiplication is a natural way of multiplying numbers: multiply the multiplicand by each digit of the multiplier and then add up all the properly shifted results.
Using long multiplication, the product of two unsigned binary integers $(a_{N-1}\cdots a_0)_{bin}$, $(b_{N-1}\cdots b_0)_{bin}$ can be calculated as
\begin{equation}
    \begin{split}
        &(a_{N-1}\cdots a_0)_{bin} \cdot (b_{N-1}\cdots b_0)_{bin}
        \\=&
        \sum_{j=0}^{N-1}2^ja_j\cdot (b_{N-1}\cdots b_0)_{bin}
    \end{split}
    \label{eq_long_mul}
\end{equation}
Thus, the multiplication of two binary numbers comes down to several binary additions of the partial products.

Hereafter we present a design of quantum multiplier based on long multiplication.

\begin{figure*}[t]
    \begin{center}
    \includegraphics[width=0.85\textwidth]{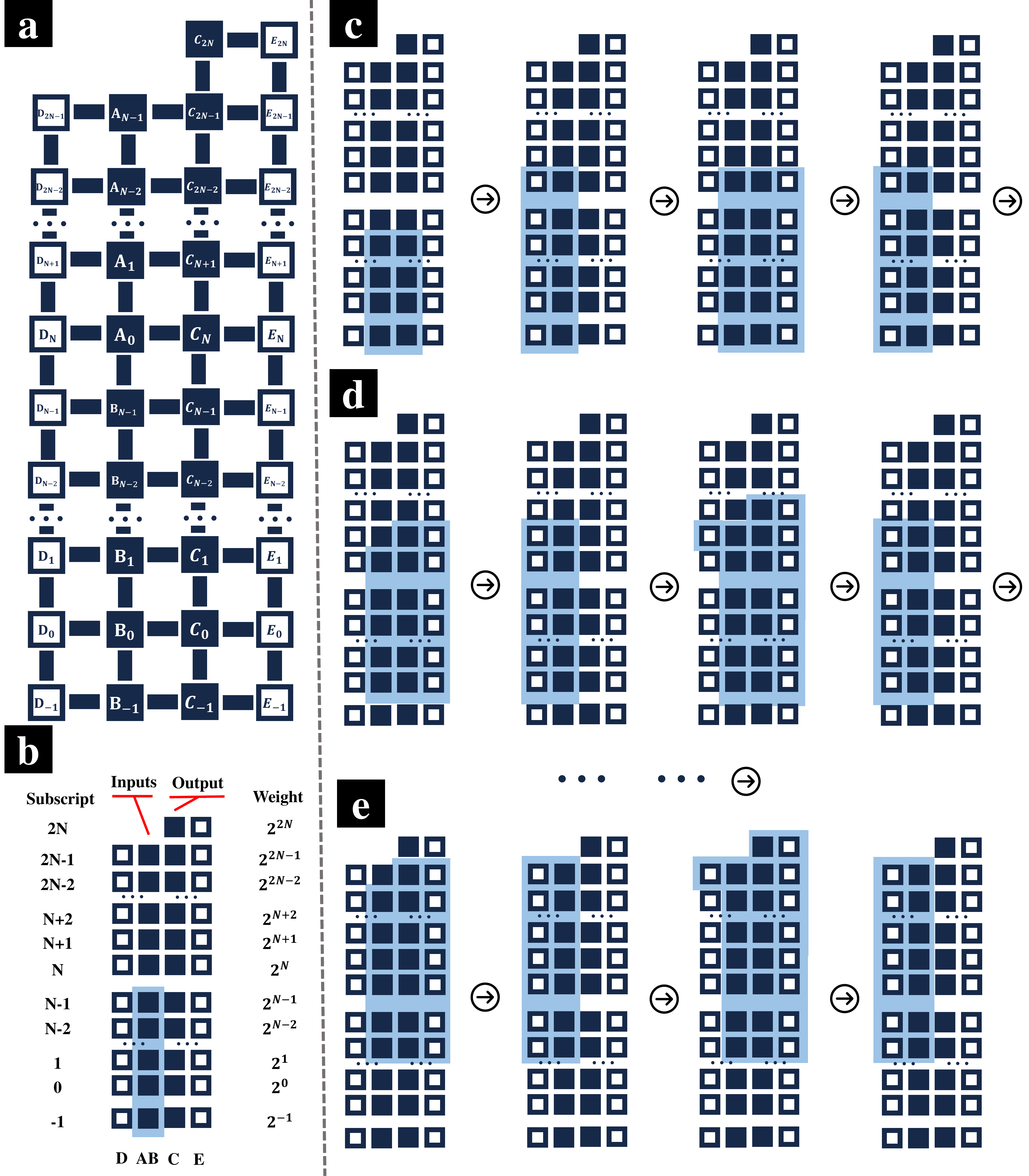}
\end{center}
    \caption{The activated qubits in quantum multiplier.
    (a) The required qubits and connectivity for quantum multiplier.
    Each square represents a qubit, and the edges indicate that the neighbor are connected physically on a quantum computer.
    The inputs are mapped to qubits $A$ and $B$ in the second column (from left), whereas the temporary sums and the final output are stored in qubits $C$ in the third column.
    Qubits $D$, $E$ in the first and last columns are ancilla qubits.
    (b,c,d,e) The activated qubits in each step, where the activated qubits are highlighted in light blue.
    (b) The activated qubits in the first step, shifting the input integer $(b_{N-1}\cdots b_0)_{bin}$ to $\frac{1}{2}(b_{N-1}\cdots b_0)_{bin}$.
    (c) The activated qubits for partial product $a_0\cdot (b_{N-1}\cdots b_0)_{bin}$.
    (d) The activated qubits for partial product $a_1\cdot (b_{N-1}\cdots b_0)_{bin}$.
    (e) The activated qubits for partial product $a_{N-1}\cdot (b_{N-1}\cdots b_0)_{bin}$.
    }
    \label{fig_mul}
\end{figure*}

The necessary qubits and their connectivity are as depicted in Fig.(\ref{fig_mul}a), where the squares represent qubits, and the edges indicate that the nearest neighbor qubits are connected physically.
There are $2N$ qubits, $A_{N-1},\cdots,A_0$ and $B_{N-1},\cdots B_0$, carrying the inputs $(a_{N-1}\cdots a_0)_{bin}$ and $(b_{N-1}\cdots b_0)_{bin}$.
Additionally, an ancilla qubit, denoted as $B_{-1}$, is required.
These $2N+1$ qubits are assigned in the second column (from left) as shown in Fig.(\ref{fig_mul}a).
Qubits $A_{N-1},\cdots,A_0$ are assigned at the top half, and then qubits $B_{N-1},\cdots B_0$ at the bottom half, from top to bottom.
The $2N+2$ qubits in the third column (from left) are denoted as $C_{2N},\cdots,C_{-1}$, from top to bottom, which are designed to store the temporary results and the final outputs.
Furthermore, $4N+3$ ancilla qubits, $D_{2N-1},\cdots,D_{-1}$ and $E_{2N},\cdots, E_{-1}$ are necessary.
As depicted in Fig.(\ref{fig_mul}a), qubits $D_{2N-1},\cdots,D_{-1}$ are assigned in the first column, and qubits $E_{2N},\cdots, E_{-1}$ in the last column (from left).
In total, there are $6N+6$ qubits.

At beginning, qubits  $A_{N-1},\cdots,A_0$ are initialized at state $|a_{N-1}\cdots a_0\rangle$, and $B_{N-1},\cdots B_0$ are initialized at $|b_{N-1}\cdots b_0\rangle$, corresponding to the input non-negative integers $(a_{N-1}\cdots a_0)_{bin}$ and $(b_{N-1}\cdots b_0)_{bin}$.
The other qubits are all prepared at state $|0\rangle$.

According to Eq.(\ref{eq_long_mul}), there are three main tasks in the long multiplication of each digit: calculating the partial product $a_j\cdot (b_{N-1}\cdots b_0)_{bin}$, shifting them, and sum up the shifted results.
For binary inputs, if $a_j=1$, the product $a_j\cdot (b_{N-1}\cdots b_0)_{bin}$ is just $(b_{N-1}\cdots b_0)_{bin}$, otherwise the product is 0. 
Intuitively, we can calculate the partial product using a quantum `Controlled Adder', where qubit $A_j$ is the control qubit.
If qubit $A_j$ is at state $|1\rangle$, the quantum adder works, summing up 
the properly shifted $(b_{N-1}\cdots b_0)_{bin}$ and the temporary results.
However, it is of great difficulty to implement the `Controlled Adder' on nowadays quantum computers (Toffoli gates among remote qubits are necessary in the implementation of such intricate operation, which can be decomposed into a series one-and two-qubit gates including considerable amounts of SWAP gates).

Here we present an alternative solution to calculate the partial product.
Notice that
\begin{equation}
    \begin{split}
        &a_j\cdot (b_{N-1}\cdots b_0)_{bin}
    \\=&
    \frac{1}{2}(b_{N-1}\cdots b_0)_{bin}
    +
    \frac{1}{2}(2a_j-1)(b_{N-1}\cdots b_0)_{bin}
    \end{split}
    \label{eq_partial}
\end{equation}
where $1/2$ indicates shifting.
The first term in Eq.(\ref{eq_partial}) is the input integer $(b_{N-1}\cdots b_0)_{bin}$.
As for the other term, due to the binary input $a_j=0,1$, we have $2a_j-1=\pm1$, which corresponds to the sign bit $\Bar{a}_j$, where $\Bar{a}_j$ is the negation of $a_j$, also denoted as $\neg a_j$.
For $a_j=0$, we have $2a_j-1=-1$, meanwhile $\Bar{a}_j=1$ (For negative integer, the sign bit is 1 in two's complement representation, as discussed in Sec.(\ref{sec_twos})).
For $a_j=1$, we have $2a_j-1=1$, and $\Bar{a}_j=0$.
In quantum computer, the Pauli-X gate conveniently converts state $|a_j\rangle$ into $|\Bar{a}_j\rangle$.
Thus, the partial product can be obtained by calculating two shifted parts, the original input $(b_{N-1}\cdots b_0)_{bin}$, and a signed integer $(\Bar{a}_j,b_{N-1}\cdots b_0)_{bin}$.
$\Bar{a}_j$ is the sign bit, $(b_{N-1}\cdots b_0)_{bin}$ is the absolute, where $\Bar{a}_j=1$ indicates negative.
Recalling the discussions in Sec.(\ref{sec_twos}), we can apply $U_{\pm}$ to prepare the two's complement representation of $(\Bar{a}_j,b_{N-1}\cdots b_0)_{bin}$, and then apply the quantum adder to sum up the temporary results.
By this mean, we can obtain the contribution of the second term in Eq.(\ref{eq_partial}) on a quantum computer.

According to Eq.(\ref{eq_long_mul}), the next task is to shift these partial products, and then sum up the temporary results.
The shifting procedure can be implemented by moving the relative states up and down.
As depicted in Fig.(\ref{fig_mul}), qubits $C_{2N},\cdots,C_{-1}$ are assigned in the third column, designed to store the temporary sums and the final output.
The subscript of qubit $C_j$ indicates that its weight is $2^j$, and qubits $A_j$(or $B_j$) and $C_j$ in one row has the same weights.
As a simple example, qubits $B_{N-1},\cdots B_0$ are prepared at state $|b_{N-1}\cdots b_0\rangle$ initially, which represents the original integer $(b_{N-1}\cdots b_0)_{bin}$, as the LSB $b_0$ is mapped to qubit $B_0$, whose weight is $2^0$.
If we shift the inputs, and set qubits $B_{N-1},\cdots B_{-1}$ at state $|b_{N-1\cdots b_0\rangle}$, then the state represents $\frac{1}{2}(b_{N-1}\cdots b_0)_{bin}$, as now the LSB $b_0$ is mapped to $B_{-1}$ and the weight of $B_{-1}$ is $2^{-1}$.
In this case, the whole state is moved down for one digit, corresponding to the shifting $2^{-1}$ (shift one bit to the right side).
Similarly, the shifting $2^N$ (shift $N$ bit to the right side) can be implemented by moving the whole quantum up for $N$ digits.
Thus, we can shift the partial product by moving the relative quantum states up and down.

We then need to sum up the shifted partial product (two terms) and the temporary sums, which can be implemented by $P_{III}$.
The shifted partial product is one input term of $P_{III}$, assigned in the second column in Fig.(\ref{fig_mul}).
Another input of $P_{III}$, the temporary sum of the previous steps, is stored in the third column in Fig.(\ref{fig_mul}).
After applying $P_{III}$, the output is stored in the corresponding $C$ qubits in the third column, which is the input temporary sum for the next digit.
For each digit, from $a_0$ to $a_{N-1}$, repeat the procedures above, and at last the product can be obtained.

In Alg.(\ref{alg_qmul}) we present the algorithm for quantum multiplier based on long multiplication.
Besides, to demonstrate the proposed quantum multiplier in a more detailed manner, we depict the activated qubits in each step as shown in Fig.(\ref{fig_mul}b,c,d,e), where the activated qubits are colored in light blue.
The tiny squares in Fig.(\ref{fig_mul}b,c,d,e) corresponds to the squares in Fig.(\ref{fig_mul}a).

The product is calculated based on long multiplication as shown in Eq.(\ref{eq_long_mul}), from LSB $a_0$ to MSB $a_{N-1}$.
After initialization, we have qubits $B_{N-1},\cdot,B_0$ at state $|b_{N-1}\cdots b_0\rangle$, corresponding to the input non-negative integer $(b_{N-1}\cdots b_0)_{bin}$.
Recalling Eq.(\ref{eq_partial}), we can break the partial product into two parts with shifting $1/2$.
Thus, it is convenient to shift the input $(b_{N-1}\cdots b_0)_{bin}$.
In Alg.(\ref{alg_qmul}), there are a series of CNOT gates in the first `while' loop, which converts qubits $B_{N-1},\cdot,B_0,B_{-1}$ to state $|0b_{N-1}\cdots b_0\rangle$.
By this mean, we shift the original input $(b_{N-1}\cdots b_0)_{bin}$ to $\frac{1}{2}(b_{N-1}\cdots b_0)_{bin}$.
In Fig.(\ref{fig_mul}b), the activated qubits are highlighted in light blue.
For clarity, we annotate the corresponding subscripts and weights in Fig.(\ref{fig_mul}b).

\begin{figure}
\begin{algorithm}[H]
\caption{Algorithm for quantum multiplier based on long multiplication}
\label{alg_qmul}
\begin{algorithmic}
\State {\bf Input:} Two N-bit binary integers: 
\State \qquad $(a_{N-1}\cdots a_0)_{bin}$, $(b_{N-1}\cdots a_0)_{bin}$
\State \qquad $A_{N-1},\cdots,A_0\gets |a_{N-1}\cdots a_0\rangle$,
\State \qquad $B_{N-1},\cdots,B_0,B_{-1}\gets |b_{N-1}\cdots b_00\rangle$,
\State \qquad $C_{2N},\cdots,C_{-1}\gets |0\cdots 0\rangle$,
\State \qquad $D_{2N-1},\cdots,D_{-1}\gets |0\cdots 0\rangle$,
\State \qquad $E_{2N},\cdots,C_{-1}\gets |0\cdots 0\rangle$.
\State (For simplicity, in this Algorithm, qubit $A_j$ is also denoted as $B_{N+j}$)

\State $n \gets 0$.
\While{$n < N$}
    \State {\bf do} SWAP $B_{n-1}$, $B_{n}$, 
    \State {\bf do} $n \gets n+1$.
    \EndWhile
    
\State $n \gets -1$.
\While{$n < N-1$}
    \State {\bf do} CNOT on $B_n$(Control), $C_n$(Target), 
    \State {\bf do} $n \gets n+1$.
    \EndWhile
\State {\bf do} Pauli-X gate on $A_0$,
\State {\bf do} $U_{\pm}$ on $A_0$(sign bit), $B_{N-1},\cdots,B_{-1}$(absolute),
\State \qquad $D_{N},\cdots,D_{-1}$(ancilla).
\State {\bf do} $P_{III}$ on $B_{N-1},\cdots,B_{-1}$(input term),
\State \qquad $C_{N},\cdots,C_{-1}$(input/output), 
\State \qquad $E_{N},\cdots,E_{-1}$(ancilla),
\State {\bf do} CNOT on $A_0$(Control), $C_N$(Target).
\State {\bf do} $U^\dagger_{\pm}$ on $A_0$(sign bit), $B_{N-1},\cdots,B_{-1}$(absolute),
\State \qquad $D_{N},\cdots,D_{-1}$(ancilla).
\State {\bf do} Pauli-X gate on $A_0$.

\State $n \gets 1$.
\While{$n < N$} 
\State $j \gets N$.
\While{$j > 0$}
    \State {\bf do} SWAP $B_{n+j-1}$, $B_{n+j-2}$, 
    \State {\bf do} $j \gets j-1$.
    \EndWhile
\State {\bf do} $P_{III}$ on $B_{N+n-1},\cdots,B_{n-1}$(input),
\State \qquad $C_{N+n},\cdots,C_{n-1}$(input/output), 
\State \qquad $E_{N+n},\cdots,E_{n-1}$(ancilla),

\State {\bf do} Pauli-X gate on $A_n$,
\State {\bf do} $U_{\pm}$ on $A_n$(sign bit), $B_{N+n-1},\cdots,B_{n-1}$, \State \qquad $D_{N+n},\cdots,D_{n-1}$(ancilla).
\State {\bf do} SWAP $A_n, D_n$,
\State {\bf do} $P_{III}$ on $B_{N+n},\cdots,B_{n-1}$(input),
\State \qquad $C_{N+n+1},\cdots,C_{n-1}$(input/output), 
\State \qquad $E_{N+n+1},\cdots,E_{n-1}$(ancilla),
\State {\bf do} SWAP $A_n, D_n$.
\State {\bf do} $U^\dagger_{pm}$ on $A_n$(sign bit), $B_{N+n-1},\cdots,B_{n-1}$,
\State \qquad $D_{N+n},\cdots,D_{n-1}$(ancilla).
\State {\bf do} CNOT on $A_n$(Control), $C_N+n$(Target).
\State {\bf do} Pauli-X gate on $A_n$.
\State $n \gets n+1$.
\EndWhile

\State $n \gets N$.
\While{$n > 0$}
    \State {\bf do} SWAP $B_{n+N-1}$, $B_{n+N-2}$, 
    \State {\bf do} $n \gets n-1$.
    \EndWhile

\State {\bf Output:} $C_{2N},\cdots,C_0\gets |m_{2N}\cdots m_0\rangle$.
\end{algorithmic}
\end{algorithm}
\end{figure}

In Fig.(\ref{fig_mul}c), the activated qubits are highlighted in light blue, which are involved in the calculation of the partial product $a_0\cdot(b_{N-1}\cdots b_0)_{bin}$.
According to Eq.(\ref{eq_partial}), the partial product is separated into two parts.
The first term is $\frac{1}{2}(b_{N-1}\cdots b_0)_{bin}$.
The input $(b_{N-1}\cdots b_0)_{bin}$ is already shifted to $\frac{1}{2}(b_{N-1}\cdots b_0)_{bin}$.
Notice that qubits $C_{2N}\cdots C_{-1}$, which store the temporary sum and the final output, is initialized at state $|0\cdots 0\rangle$.
Thus, we do not need to apply the full $P_{III}$ to add up $\frac{1}{2}(b_{N-1}\cdots b_0)_{bin}$ and 0.
Instead, we can apply a series of CNOT gates, where qubits $B_{N-2}\cdots B_{-1}$ are control qubits, and $C_{N-2}\cdots C_{-1}$ are targets.
In the first column of Fig.(\ref{fig_mul}c), the activated qubits in the CNOT series are highlighted.
Then consider the other term, $\frac{1}{2}(2a_0-1)(b_{N-1}\cdots b_0)_{bin}$, or equivalently, $-\frac{1}{2}\Bar{a}_0\cdot(b_{N-1}\cdots b_0)_{bin}$.
A Pauli-X gate converts qubit $A_0$ to state $|\Bar{a}_0\rangle$.
Next, $U_{\pm}$ is applied on qubits $A_0,B_{N-1},\cdots,B_{-1}$ and $D_{N},\cdots,D_{-1}$, where $A_0$ corresponds to the sign bit, $B_{N-1},\cdots,B_{-1}$ correspond to the absolute, and $D_{N},\cdots,D_{-1}$ are ancilla qubits.
By this mean, we get the two's complement representation of signed binary integer $(\Bar{a}_0,0b_{N-1}\cdots b_{0})_{bin}$, where $\Bar{a}_0$ is the sign bit.
Notice that we have shifted the input, there is an additional 0 (The additional 0 avoids the possible overflow).
If $a_0=1$, then $\Bar{a}_0=0$, and $U_{\pm}$ does not change anything.
Otherwise, $a_0=0$, $\Bar{a}_0=1$, and $U_{\pm}$ converts qubits $B_{N-1},\cdots,B_{-1}$ to the appropriate two's complement representation, corresponding to a negative integer.
Then apply $P_{III}$ on qubits $B_{N-1},\cdots,B_{-1}$, $C_{N},\cdots,C_{-1}$ and $E_{N},\cdots,E_{-1}$, where qubits $E_{N},\cdots,E_{-1}$ are ancilla.
Notice that the sign bit is not included in this step.
At this stage, qubits $C_N, C_{N-1},\cdots,C_{-1}$ are at state $|00b_{N-1}\cdots b_0\rangle$.
$P_{III}$ adds the temporary results up.
If $a_0=1$, $\Bar{a}_0=0$, we will have qubits $C_N, C_{N-1},\cdots,C_{-1}$ at state $|0b_{N-1}\cdots b_00\rangle$, corresponding to $(b_{N-1}\cdots b_{0})_{bin}$.
Otherwise, $a_0=0$, $\Bar{a}_0=1$, we will have qubits $C_N, C_{N-1},\cdots,C_{-1}$ at state $|10\cdots 0\rangle$, where all digits are 0 except $C_N$.
This result indicates that the temporary sum is 0, and the MSB is 1 as overflow.
Till now the sign bit $A_0$ is excluded in the addition.
To take the sign bit into account, we then apply a CNOT gate between qubit $A_0$ and $C_N$, where $A_0$ is control qubit.
For $a_0=1$, $\Bar{a}_0=0$, the CNOT gate does not change anything.
Whereas for $a_0=0$, $\Bar{a}_0=1$, we will have qubit $C_N$ converted to state $|0\rangle$.
Notice that qubits $A_0,B_{N-1},\cdots,B_{-1}$ are still at the quantum state corresponding to the two's complement representation.
We need to reset these qubits.
Thus, $U^\dagger_{\pm}$ is applied on qubits $A_0,B_{N-1},\cdots,B_{-1}$.
Then a Pauli-X gate is applied on qubit $A_0$, resetting it to the original state $|a_0\rangle$.
At end, a series of CNOT gates are applied, shifting up $(b_{N-1}\cdots b_0)_{bin}$ for one digit.

Then consider the partial product $a_1\cdot(b_{N-1}\cdots b_0)_{bin}$.
In Fig.(\ref{fig_mul}d), the relative activated qubits are highlighted in light blue.
We have already shifted the quantum state properly.
Qubits $A_1, A_0,B_{N-1},\cdots,B_{-1}$ are at state $|a_10b_{N-1}\cdots b_0 a_0\rangle$.
Still, the partial product can be separated into two parts, $\frac{1}{2}(b_{N-1}\cdots b_0)_{bin}$, and $-\frac{1}{2}\Bar{a}_1\cdot(b_{N-1}\cdots b_0)_{bin}$.
To add up the first part and the temporary sum, $P_{III}$ is applied on qubits $A_0,B_{N-1},\cdots,B_0$, $C_{N+1},\cdots,C_0$, and ancilla qubits $E_{N+1},\cdots,E_0$.
Then consider the second term.
A Pauli-X gate converts the sign bit $A_1$ to state $|\Bar{a}_1\rangle$, and $U_{\pm}$ is applied to prepare the two's complement representation.
Additionally, a SWAP gate is applied, swapping qubits $A_1$ and $D_1$.
By this mean, we have qubits $A_1$ at state $|0\rangle$.
At this stage, the temporary sum contains $2^0\cdot a_0\cdot(b_{N-1}\cdots b_0)_{bin}$, and $\frac{1}{2}\cdot2^1\cdot a_1\cdot(b_{N-1}\cdots b_0)_{bin}$.
Thus, $C_{N+1}$ corresponds to the first non-trivial digit in the temporary sum($C_{N+2},\cdots,C_{2N}$ are all at state $|0\rangle$ now).
Then $P_{III}$ is applied on qubits $A_1, A_0, B_{N-1},\cdots,B_0$, $C_{N+2},\cdots,C_0$, and ancilla qubits $E_{N+2},\cdots,E_0$.
After the addition, the SWAP gate is applied again to swap qubits $A_1$ and $D_1$.
Next, a CNOT gate is applied, where $A_1$ is the control qubit, and $C_{N+}1$ is the target.
If $a_1=1$, $\Bar{a}_1=0$, then $\frac{1}{2}\cdot2^1\cdot a_1\cdot(b_{N-1}\cdots b_0)_{bin}$ is added to the temporary sum, and the new temporary sum can be given as $(a_1a_0)_{bin}\cdot (b_{N-1}\cdots b_0)_{bin}$.
The product of a two-digit binary number and a $N$-digit binary number is no greater than $2^{2+N}-1$, and can be described by a $N+2$ digit binary number.
Thus, the temporary sum is stored in qubits $C_{N+1},\cdots, C_0$, without any overflow.
As $\Bar{a}_1=0$, the succeeding CNOT gate does not change anything.
Otherwise, $a_1=0$, $\Bar{a}_1=1$, then the two's complement representation of $(b_{N-1}\cdots b_0)_{bin}$ is included.
Recalling that the summation of $(0b_{N-1}\cdots b_0)_{bin}$ and the two's complement representation of its negative is $2^{N+1}$, corresponding to an overflow to the higher order digit.
Thus, for $a_1=0$, $\Bar{a}_1=1$, we have the new temporary sum is 
$$
a_0\cdot(b_{N-1}\cdots b_0)_{bin}+2^{N+1}
$$
where $2^{N+1}$ indicates that qubit $C_{N+1}$ is at state $|1\rangle$.
As $\Bar{a}_1=1$, the succeeding CNOT gate reset qubit $C_{N+1}$ to state $|0\rangle$.
By this mean, contributions of the product $a_1\cdot(b_{N-1}\cdots b_0)_{bin}$ cancel out.
By the end, we need to reset the relative qubits, and shift the state properly.
$U_{\pm}$ and Pauli-X gates are applied again to reset the relative qubits, and a series of CNOT gates are applied for shifting.

The iterative procedures above are repeated for each digit.
In Fig.(\ref{fig_mul}e) we present the activated qubits in the calculation of the last partial product $a_{N-1}\cdot(b_{N-1}\cdots b_0)_{bin}$.
The first column (from left) in Fig.(\ref{fig_mul}c,d,e) corresponds to the addition of the first term of the partial product.
In the second column we highlighted the activated qubits in the preparation of the two's complement representation.
Next, the third column corresponds to the addition of the second term of the partial product.
In the last column the highlighted qubits are activated for reset and shifting.

After applying the quantum multiplier, we have qubits $C_{2N},\cdots,C_0$ at state $|m_{2N}\cdots m_0\rangle$, where $(m_{2N}\cdots m_0)_{bin}$ is the product of the inputs.
The ancilla qubits $D$ and $E$ are all reset to state $|0\rangle$.
Meanwhile, the quantum multiplier `swap' the two inputs in the iterative shifting process.
By the end, we have qubits $B_{N-2},\cdots, B_{-1}$ at state $|a_{N-1}\cdots a_0\rangle$, whereas qubits $A_{N-1},A_{N-2},\cdots, A_{0},B_{N-1}$ at state $|0b_{N-1}\cdots b_1b_0\rangle$.
Thus, if the inputs are used in succeeding operations, we still need to reset them by repeating the shifting process.

Then study the time complexity of the proposed quantum multiplier.
For each digit, there are four main tasks: shifting, adding up the first term of the partial product, preparing the two's complement representation, adding the second term, and reset.
To shift the quantum state up or down for one digit, $N$ CNOT gates are required.
Next, the time complexity of quantum adder $P_{III}$ is of the order $\mathcal{O}(N)$.
As for the two's complement representation, the time complexity of $U_{\pm}$ is also $\mathcal{O}(N)$.
The inputs are both $N$-digit binary integers, and the tasks above are repeated for $N$ times.
Therefore, the overall time complexity of the proposed quantum multiplier is $\mathcal{O}(N^2)$.
Similar to $P_{III}$ and $U_{\pm}$, the proposed quantum multiplier can be implemented using only Pauli-X gates, CNOT gates, and $C\sqrt{X}$ (CSX) gates, and all two-qubit gates are operated between nearest neighbor qubits.

\section{Quantum Divider}
\label{sec_divider}

\subsection{Quantum Divider Based on Long Division}
\label{sec_divider1}

Like long multiplication, long division is a standard division algorithm.
The first step of long division is to divide the left-most digit of the  numerator (dividend) by the  denominator (divisor). 
The quotient (rounded down to an integer) becomes the first digit of the result, and the remainder is then calculated by subtraction.
This remainder is then carried forward to the next digit of the dividend, where the division process is reiterated.
When all digits have been processed and no remainder is left, the process is complete.

In this section we present a quantum binary divider based on long division algorithm.
For given numerator (dividend) $(a_{N-1}\cdots a_0)_{bin}$ and denominator (divisor) $(b_{M-1}\cdots b_0)_{bin}$, our aim is to calculate the quotient $({q_{N-1}\cdots q_0})_{bin}$ and remainder $({r_{M-1}\cdots r_0})_{bin}$.
The numerator (dividend) and quotient are $N$-digit binary integers, whereas the denominator (divisor) and remainder are $M$-bit binary integers.
For simplicity, we can set $M\leq N$ (In fact, even if $M>N$, we can lengthen the numerator by placing trivial zeros as the first $M-N$ digits).
Meanwhile, $b_{M-1}$ can be either 0 or 1.
In other words, even if there are trivial zeros in the given denominator, the proposed quantum divider can still work.

\begin{figure*}[t]
    \begin{center}
    \includegraphics[width=0.95\textwidth]{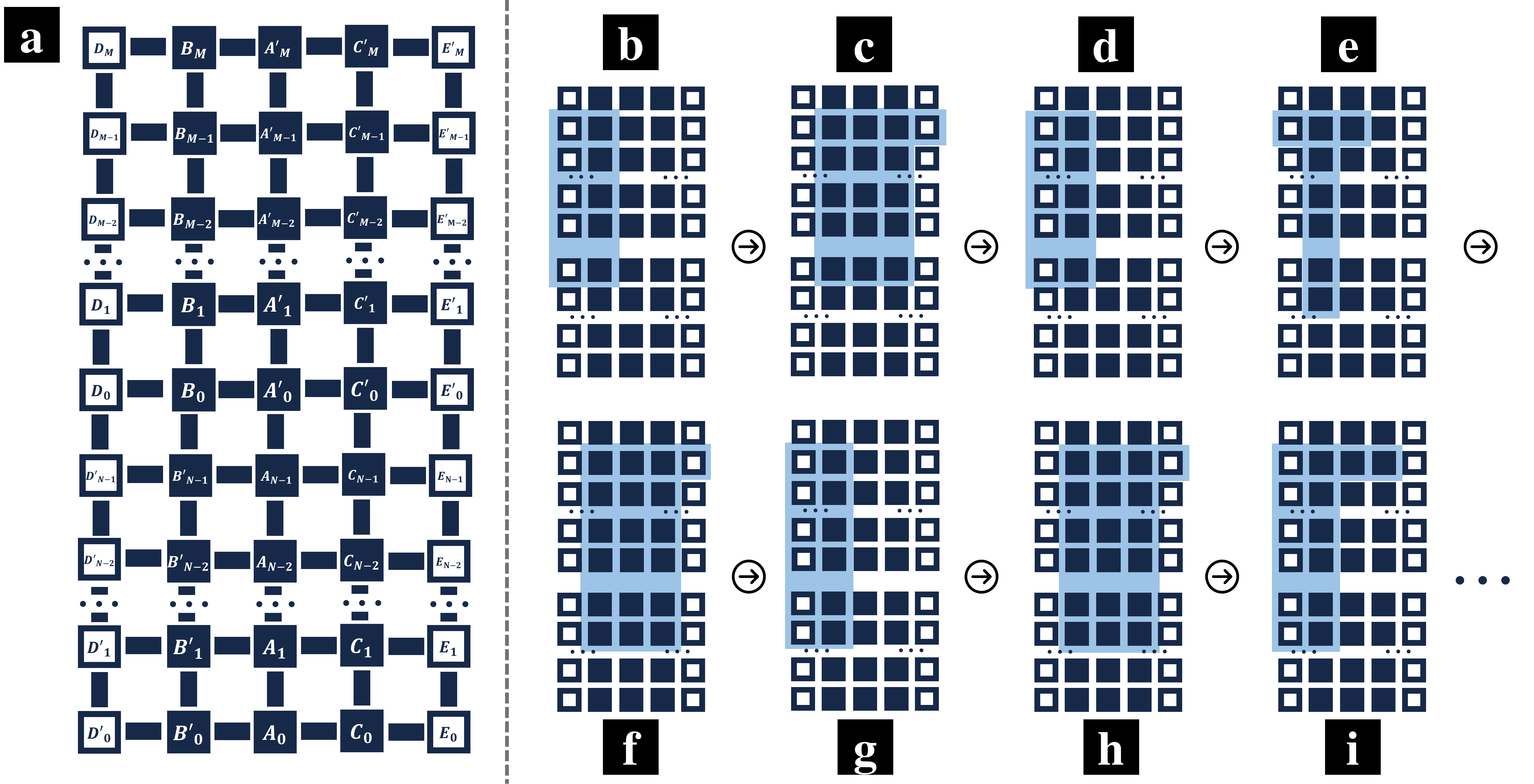}
\end{center}
    \caption{The diagram of connectivity and activated qubits in quantum divider.
    (a) The required qubits and connectivity for quantum divider.
    Each square represents a qubit, and the edges indicate that the neighbors are connected physically on a quantum computer.
    The inputs are mapped to qubits $A$ and $B$ in the second column (from left), whereas the temporary sums and the final output are stored in qubits $C$ in the third column.
    Qubits $D$, $E$ in the first and last columns are ancilla qubits.
    (b-i) The activated qubits involved in the division for the first digit of the dividend (MSB $a_{N-1}$). 
    }
    \label{fig_divider}
\end{figure*}

Fig.(\ref{fig_divider}a) is a diagram of the required qubits and connectivity for the proposed quantum divider, where each square represents a qubit, and the edges indicate that the nearest neighbor qubits are connected physically.
There are in total $5(N+M+1)$ qubits.
As depicted in Fig.(\ref{fig_divider}a), these qubits are assigned in 5 columns, denoted as $D,D';B,B';A',A;C',C;E',E$, from left to right.
In $D,D'$ and $B,B'$ columns, the qubits in the top part are denoted as $D$ or $B$, and qubits in the bottom part are denoted as $D'$ or $B'$.
On the contrary, in the other three columns, qubit in the top part are denoted as $A'$, $C'$ or $E'$, whereas qubits in the bottom part are denoted as $A$, $C$, $E$.
Qubits in the same row share the same subscripts, which indicate their weights in the calculation.
Additionally, to implement the full divider, 4 additional qubits $D'_{-1}$, $B'_{-1}$, $A_{-1}$ and $C_{-1}$ are required, which are assigned in one row, connected to qubits $D'_{0}$, $B'_{0}$, $A_{0}$ and $C_{0}$.
If we need the output reminder, then these 4 qubits are necessary.
Otherwise, if we only need the output quotient, then we can discard these 4 qubits.
Thus for simplicity, we do not plot these 4 qubits in Fig.(\ref{fig_divider}).

In brief, the proposed quantum divider works as follows.

\noindent
{\bf 0.} Initialization. 
The input numerator (dividend) $(a_{N-1}\cdots a_0)_{bin}$ is mapped to $A',A$ column, and the denominator (divisor) $(b_{M-1}\cdots b_0)_{bin}$ is mapped to $B, B'$ column.
Align the LSB of the divisor, $b_0$, and the MSB of the dividend $a_{N-1}$ in the same row.

\noindent
{\bf 1.} Calculate the two's complement representation of $-(b_{M-1}\cdots b_0)_{bin}$.

\noindent
{\bf 2.} The division starts from the MSB $a_{N-1}$. 
Calculate the summation of $a_{N-1}$ and $-(b_{M-1}\cdots b_0)_{bin}$.
Denote the sign bit of the summation is $\Bar{c}_{N-1}$.

\noindent
{\bf3.} If the summation is negative, $\Bar{c}_{N-1}=1$, then add up $(b_{M-1}\cdots b_0)_{bin}$ and the temporary sum.

\noindent
{\bf4.}
If $a_{N-1}$ is not the LSB, then shift the divisor for one digit, aligning $b_0$ and $a_{N-2}$ in the same row.

\noindent
{\bf5.} Repeat 1-4 until all digits have been processed.
The quotient is $({c}_{N-1}\cdots {c}_{0})_{bin}$, and the dividend has been converted to the remainder $(d_{a_{N-1}\cdots d_0})_{bin}$ in the iterative subtraction and addition.

Hereafter we present the detailed implementation for each step.
At beginning, the given numerator (dividend) $(a_{N-1}\cdots a_0)_{bin}$ is mapped to qubits $A_{N-1},\cdots,A_0$, and denominator (divisor) $(b_{M-1}\cdots b_0)_{bin}$ is mapped to qubits $B_{M-1},\cdots,B_0$.
For the convenience to implement long division, qubits $A_{N-1},\cdots,A_0$ are assigned in the bottom part of the $A', A$ column, whereas qubits qubits $B_{M-1},\cdots,B_0$ are assigned at the top part of the $B, B'$ column.
All the other qubits are initialized at state $|0\rangle$.
Qubits $D,D'$ and $E',E$ are all ancilla qubits, which will be reset at state $|0\rangle$ after applying the full quantum divider.
Qubits $C,C'$ acts like ancilla qubits, yet the output quotient will be stored in these qubits by the end.
Notice that $b_0$ and $a_{N-1}$ are not aligned in the same row yet.
To shift the input divisor $(b_{M-1}\cdots b_0)_{bin}$ for one digit, we still need to apply a series of SWAP gates.
After the shifting process, we have qubits $B_{M-1}, B_{M-2},\cdots,B_0, B'_{N-1}$ at state $|0b_{M-1}\cdots b_1b_0\rangle$.

The next step is to calculate the two's complement representation of $-(b_{M-1}\cdots b_0)_{bin}$.
Recalling Sec.(\ref{sec_twos}), operation $U_{\pm}$ can convert the given signed integer to its two's complement representation.
Notice that we have shifted the original input  $(b_{M-1}\cdots b_0)_{bin}$ for one digit, and qubit $B_{M-1}$ is at state $|0\rangle$ after the shifting process.
Thus, we can set qubit $B_{M-1}$ as the sign bit.
A Pauli-X gate flips qubit $B_{M-1}$ to state $|1\rangle$, indicating that the signed integer is negative.
Next, $U_{\pm}$ is applied on qubits $B_{M-1}, B_{M-2},\cdots,B_0, B'_{N-1}$ and ancilla qubits $D_{M-1}, D_{M-2},\cdots,D_0, D'_{N-1}$.
By this mean, we have prepared the two's complement representation of $-(b_{M-1}\cdots b_0)_{bin}$.
In Fig.(\ref{fig_divider}b) the activated qubits in this step are highlighted with light blue.

Then in step 2, the subtraction $(0\cdots a_{N-1})_{bin}-(b_{M-1}\cdots b_0)_{bin}$ is calculated.
As we have already prepared the two's complement representation of $-(b_{M-1}\cdots b_0)_{bin}$, here we alternatively calculate the summation of $-(b_{M-1}\cdots b_0)_{bin}$ and $(0\cdots a_{N-1})_{bin}$.
The summation of these two signed integers can be obtained by applying $\Tilde{P}_{III}$, as discussed in Sec.(\ref{sec_subtractor}). 
Operation $\Tilde{P}_{III}$ then acts on qubits $B_{M-1}, B_{M-2},\cdots,B_0, B'_{N-1}$ (input), $A'_{M-1}, A'_{M-2},\cdots,A'_0, A_{N-1}$ (input and output), and $E'_{M-1},C'_{M-1}, C'_{M-2},\cdots,C'_0, C_{N-1}$ (ancilla).
The temporary result is store in qubits $A'_{M-1}, A'_{M-2},\cdots,A'_0, A_{N-1}$ in two's complement representation, where qubit $A'_{M-1}$ is the sign bit.
Denote the sign bit of the temporary result as $\Bar{c}_{N-1}$.
If $(b_{M-1}\cdots b_0)_{bin} > (0\cdots a_{N-1})_{bin}$, the temporary result is negative, and we have $\Bar{c}_{N-1}=1$.
Otherwise $(b_{M-1}\cdots b_0)_{bin} \leq (0\cdots a_{N-1})_{bin}$, and the temporary result is non-negative, we have $\Bar{c}_{N-1}=0$.
According to long division, for $(b_{M-1}\cdots b_0)_{bin} > (0\cdots a_{N-1})_{bin}$, the first digit of the quotient is 0, otherwise 1.
Therefore, we have the $c_{N-1}$ as the first digit of the quotient.

In long division, we do the subtraction only if the first digit of the quotient is 1, thus the reminder is always non-negative.
However, in the proposed quantum divider, we always calculate the subtraction, and the reminder can be either positive or not.
Therefore in step 3, our aim is to add up $\Bar{c}_{N-1}\cdot(b_{M-1}\cdots b_0)_{bin}$ and the temporary sum.
Recalling Eq.(\ref{eq_partial}), we have
\begin{equation}
    \begin{split}
        &\Bar{c}_{N-1}\cdot(b_{M-1}\cdots b_0)_{bin}
        \\
        =&
        \frac{1}{2}(b_{M-1}\cdots b_0)_{bin}
        +\frac{1}{2}(2\Bar{c}_{N-1}-1)(b_{M-1}\cdots b_0)_{bin}
    \end{split}
    \label{eq_partial_divider}
\end{equation}
For convenience, in the quantum implementation we add up two alternative terms along with the temporary sum, $\frac{1}{2}(b_{M-1}\cdots b_0)_{bin}$ and $\frac{1}{2}(2\Bar{c}_{N-1}-1)(b_{M-1}\cdots b_0)_{bin}$.

Consider the first term in Eq.(\ref{eq_partial_divider}), $\frac{1}{2}(b_{M-1}\cdots b_0)_{bin}$.
By the end of step 2, the original input $(b_{M-1}\cdots b_0)_{bin}$ has been converted to the two's complement representation of $(-b_{M-1}\cdots b_0)_{bin}$.
Thus, $U_{\pm}^\dagger$ is applied to qubits $B_{M-1}, B_{M-2},\cdots,B_0, B'_{N-1}$ and ancilla qubits $D_{M-1}, D_{M-2},\cdots,D_0, D'_{N-1}$, resetting qubits $B_{M-1}, B_{M-2},\cdots,B_0, B'_{N-1}$ to state $|1b_{M-1}\cdots b_1 b_0\rangle$.
Then a Pauli-X gate flips qubit $B_{M-1}$, and the sign bit is flipped from 1 to 0.
In Fig.(\ref{fig_divider}d), the activated qubits in the resetting process are highlighted.
Next, the divisor is shifted for one digit, corresponding to the amplitude $\frac{1}{2}$ in Eq.(\ref{eq_partial_divider}).
Notice that the succeeding summation might change the sign bit of the temporary sum stored in the $A',A$ column, we still need to make a `copy' of $\Bar{c}_{N-1}$.
Thus, four CNOT gates are applied, ancilla qubit $D_{N-1}$ is set as state $|\Bar{c}_{N-1}\rangle$.
The activated qubits in the above process are highlighted in Fig.(\ref{fig_divider}e).
Afterward, sum up the shifted divisor and the temporary sum by applying $\Tilde{P}_{III}$, as shown in Fig.(\ref{fig_divider}f).
The temporary sum now is the two's complement representation of 
\begin{equation*}
    \begin{split}
    &(0\cdots a_{N-1}a_{N-2})_{bin}-(b_{M-1}\cdots b_00)_{bin}
    \\
    &+\frac{1}{2}(b_{M-1}\cdots b_00)_{bin}
    \\
    =&(0\cdots a_{N-1}a_{N-2})_{bin}-(b_{M-1}\cdots b_0)_{bin}
\end{split}
\end{equation*}
where the additional 0 indicates that we have shifted the divisor for one digit, and we have $\frac{1}{2}(b_{M-1}\cdots b_00)_{bin}=(b_{M-1}\cdots b_0)_{bin}$.
Meanwhile, qubit $A'_{M-1}$ stores the information of the sign bit for the temporary sum.

Then consider the contribution of the second term in Eq.(\ref{eq_partial_divider}), $\frac{1}{2}(2\Bar{c}_{N-1}-1)(b_{M-1}\cdots b_0)_{bin}$.
If $\Bar{c}_{N-1}=1$, $2\Bar{c}_{N-1}-1=1$, then our aim is to add up $\frac{1}{2}(b_{M-1}\cdots b_0)_{bin}$.
Otherwise, $\Bar{c}_{N-1}=0$, $2\Bar{c}_{N-1}-1=-1$, then we need to add up $-\frac{1}{2}(b_{M-1}\cdots b_0)_{bin}$.
As discussed in Sec.(\ref{sec_multiplier}), we can treat the second term as a signed integer, and $2\Bar{c}_{N-1}-1$ corresponds to the sign bit $c_{N-1}$.
After adding up the first term in Eq.(\ref{eq_partial_divider}), we have qubit $B_{M-1}$ at state $|0\rangle$, and ancilla qubit $D_{M-1}$ at state $|\Bar{c}_{N-1}\rangle$.
Thus, we can swap qubits $B_{M-1}$ and $D_{M-1}$, then apply Pauli-X gate on qubit $B_{M-1}$, flipping it from state $|\Bar{c}_{N-1}\rangle$ to $|{c}_{N-1}\rangle$.
Next, $U_{\pm}$ is applied on qubits $B_{M-1}, B_{M-2},\cdots,B_0, B'_{N-1}$ and ancilla qubits $D_{M-1}, D_{M-2},\cdots,D_0, D'_{N-1}$, and the activated qubits are highlighted in Fig.(\ref{fig_divider}g).
By this mean, we have qubits $B_{M-1}, B_{M-2},\cdots,B_0, B'_{N-1}$ at the two's complement representation, where qubit $B_{M-1}$ is at state $|{c}_{N-1}\rangle$, corresponding to the sign bit $c_{N-1}$.
Afterward, $\Tilde{P}_{III}$ is applied, adding up the second term and the temporary result, and the activated qubits are highlighted in Fig.(\ref{fig_divider}h).

To add up the two terms in Eq.(\ref{eq_partial_divider}) along with the temporary result, there is one `unexpected' qubit $A_{N-2}$ involved in the operation $\Tilde{P}_{III}$, as depicted in (\ref{fig_divider}f,h).
If $c_{N-1}=0$, then the total contribution of these two terms is $(b_{M-1}\cdots b_0)_{bin}$, where the LSB $b_0$ is of the same weight as $a_{N-1}$.
In this case, qubit $A_{N-2}$ is still at state $|a_{N-2}\rangle$ after the summations.
Otherwise $c_{N-1}=1$, then the total contribution of these two terms cancels out.
Therefore, the additions do not change the state of qubit $A_{N-2}$.

Then study the temporary summation.
For $\Bar{c}_{N-1}=1$, $c_{N-1}=0$, we have $(0\cdots a_{N-1})_{bin}<(b_{M-1}\cdots b_0)_{bin}$.
In this case, we add up $(b_{M-1}\cdots b_0)_{bin}$ and the temporary summation, resetting the temporary summation to $(0\cdots a_{N-1})_{bin}$.
$(0\cdots a_{N-1})_{bin}$ is the reminder for the division of next digit, and the first digit of the quotient is obtained as $c_{N-1}=0$.
On the contrary, if $\Bar{c}_{N-1}=0$, $c_{N-1}=1$, we have $(0\cdots a_{N-1})_{bin}\geq(b_{M-1}\cdots b_0)_{bin}$.
In this case, the two terms in Eq.(\ref{eq_partial_divider}) cancel out, and the temporary summation is $(0\cdots a_{N-1})_{bin}-(b_{M-1}\cdots b_0)_{bin}\geq0$.
Meanwhile, we have first digit of the quotient is $c_{N-1}=1$.
Till now, we have implemented step 3 and 4.
(Notice if $a_{M-1}$ is the last digit of the dividend, and we do not need the output reminder, then step 3 and 4 are not necessary. In this case, the qubits $D'_{-1}$, $B'_{-1}$, $A_{-1}$ and $C_{-1}$ are no more necessary.)

By the end, we need to reset the divisor to $(b_{M-1}\cdots b_0)_{bin}$.
Thus, $U_{\pm}^\dagger$ is applied on qubits $B_{M-1}, B_{M-2},\cdots,B_0, B'_{N-1}$ and ancilla qubits $D_{M-1}, D_{M-2},\cdots,D_0, D'_{N-1}$.
Besides, SWAP gates are applied on qubits $B_{M-1}$, $A'_{M-1}$, $C'_{M-1}$, resetting qubit $B_{M-1}$ to state $|0\rangle$, and converting qubit $C'_{M-1}$ to state $|c_{N-1}\rangle$.
Thus, the output quotient is stored in the $C',C$ column.

In Alg.(\ref{alg_divider}) we present the algorithm for quantum divider based on long division.
For simplicity, in Alg.(\ref{alg_divider}) qubits $B_{j}$, $D_{j}$ are also denoted as $B'_{j+N}$, $D'_{j+N}$, and qubits $A'_{j}$, $C'_{j}$, $E'_{j}$ are also denoted as $A_{j+N}$, $C_{j+N}$, $E_{j+N}$.
After applying the quantum divider, qubits $C_{N+M-1},\cdots,C_{M}$ are converted to quantum state $|q_{N-1}\cdots q_0\rangle$, corresponding to the quotient, a $N$ bit integer denoted as $(q_{N-1}\cdots q_0)_{bin}$.
Qubits $A_{M-1},\cdots,A_{0}$ are at state $|r_{M-1}\cdots r_0\rangle$, corresponding to the reminder, a $M$ bit integer $(r_{M-1}\cdots r_0)_{bin}$.
Meanwhile, the input divisor is `shifted', and we have qubits $B'_{M-2},\cdots,B'_{-1}$ at state $|b_{M-1}\cdots b_0\rangle$.
Other qubits in column $A',A$, $B,B'$ and $C',C$, along with all ancilla qubits in column $D,D'$ and $E',E$ are at state $|0\rangle$.

\begin{figure}
\begin{algorithm}[H]
\caption{Algorithm for Quantum Divider Based on Long Division}
\label{alg_divider}
\begin{algorithmic}
\State {\bf Input:} Two non-negative binary integers: 
\State \qquad Dividend: $A_{N-1},\cdots,A_0\gets (a_{N-1}\cdots a_0)_{bin}$,
\State \qquad Divisor: $B_{M-1},\cdots,B_0\gets (b_{M-1} \cdots b_0)_{bin}$,
\State \qquad Other qubits are all initialized at state $|0\rangle$.
\State (For simplicity, in this Algorithm, 
qubits $B_{j}$, $D_{j}$ are also denoted as $B'_{j+N}$, $D'_{j+N}$, 
qubits $A'_{j}$, $C'_{j}$, $E'_{j}$ are also denoted as $A_{j+N}$, $C_{j+N}$, $E_{j+N}$.)
\State $j \gets 0$
\While{$j < M$}
    \State {\bf do} SWAP on $B'_{j+N}$, $B'_{j+N-1}$.
    \State {\bf do} $j \gets j+1$.
    \EndWhile
    
\State $n \gets N-1$    
\While{$n \geq 0$}
    \State {\bf do} Pauli-X on $B'_{M+n}$,
    \State {\bf do} $U_{\pm}$ on $B'_{M+n},\cdots, B'_{n}; D'_{M+n}, \cdots, D'_{n}$.
    \State {\bf do} $\Tilde{P}_{III}$ on $B'_{M+n},\cdots, B'_{n}; A_{M+n}, \cdots, A_{n};$
    \State \qquad $ C_{M+n}, \cdots, C_{n}; E_{M+n}$.
    \State {\bf do} $U^\dagger_{\pm}$ on $B'_{M+n},\cdots, B'_{n}; D'_{M+n}, \cdots, D'_{n}$,
    \State {\bf do} Pauli-X on $B'_{M+n}$.
    
    \State {\bf do} SWAP on $A_{n+N}$, $B'_{n+N}$,
    \State {\bf do} SWAP on $B'_{n+N}$, $D'_{n+N}$.
    
    \State $j \gets 0$,
    \While{$j < M$}
        \State {\bf do} SWAP on $B'_{n+j}$, $B'_{n+j-1}$.
        \State {\bf do} $j \gets j+1$.
        \EndWhile

    \State {\bf do} $\Tilde{P}_{III}$ on $B'_{M+n},\cdots, B'_{n-1}; A_{M+n}, \cdots, A_{n-1};$
    \State \qquad $ C_{M+n}, \cdots, C_{n-1}; E_{M+n}$.
    
    \State {\bf do} SWAP on $B'_{n+N}$, $D'_{n+N}$,
    \State {\bf do} Pauli-X on $B'_{M+n}$.
    \State {\bf do} $U_{\pm}$ on $B'_{M+n},\cdots, B'_{n-1}; D'_{M+n}, \cdots, D'_{n-1}$.

    \State {\bf do} $\Tilde{P}_{III}$ on $B'_{M+n},\cdots, B'_{n-1}; A_{M+n}, \cdots, A_{n-1};$
    \State \qquad $ C_{M+n}, \cdots, C_{n-1}; E_{M+n}$.

    \State {\bf do} $U^\dagger_{\pm}$ on $B'_{M+n},\cdots, B'_{n-1}; D'_{M+n}, \cdots, D'_{n-1}$.
    \State {\bf do} SWAP on $B'_{n+N}$, $A_{n+N}$,
    \State {\bf do} SWAP on $A_{n+N}$, $C_{n+N}$.
    
    \State {\bf do} $n \gets n-1$.
    \EndWhile

\State {\bf Output:} 
\State \qquad Shifted divisor: $B'_{M-2},\cdots,B'_{-1}\gets |b_{M-1}\cdots b_0\rangle$,
\State \qquad Quotient: $C_{N+M-1},\cdots,C_{M}\gets |q_{N-1}\cdots q_0\rangle$,
\State \qquad Reminder: $A_{M-1},\cdots,A_{0}\gets |r_{M-1}\cdots r_0\rangle$,
\end{algorithmic}
\end{algorithm}
\end{figure}

At the end of this subsection we would like to give a brief discussion about the time complexity of Alg.(\ref{alg_divider}).
In step 1-4, we implement the long division for one single digit of the dividend (specifically, the MSB).
We need to repeat these steps for $N$ times, until all digits have been processed.
In step 1-4, $U_{\pm}$ (or $U^\dagger_{pm}$) is applied for four times, $\Tilde{P}_{III}$ is applied for three times, and we have shifted the divisor for one digit. 
In total, the time complexity of the proposed quantum divider is of the order $\mathcal{O}(NM)$.
As $M\leq N$, the time complexity is of the order $\mathcal{O}(N^2)$.

\subsection{Division by Zero}
In Sec.(\ref{sec_divider1}), both the divisor and the dividend are supposed to be non-negative integers.
For inputs without superposition, it is feasible to eliminate the zero divisors.
However, the proposed quantum binary divider also supports input with superposition, and we can hardly avoid the zero divisors in the superposition.
Thus, we have to face the problematic special case, division by zero, where the divisor (denominator) is zero.
Mathematically, the quotient tends to be infinity in this case. 
Yet in the proposed quantum divisor, the output can never be `infinity' rigorously.

Recalling Alg.(\ref{alg_divider}), when the input divisor is 0, the summation of $a_{N-1}$ and $-(b_{M-1}\cdots b_0)_{bin}$ is just $a_{N-1}$, which is non-negative.
Therefore, the first digit of the output quotient is always 1.
Then consider the adding back process, according to Eq.(\ref{eq_partial_divider}).
As the divisor $(b_{M-1}\cdots b_0)_{bin}$ is 0, both the two terms in Eq.(\ref{eq_partial_divider}) lead to 0 (the two's complement representation for $\pm 0$ is always a string of zeros).
Thus, the adding back process does not change the reminder.
For positive divisors, each time we do step 1-3, it is guaranteed that the reminder is no greater than the reminder is less than the divisor, and then we can safely shift the divisor, repeating the calculation for next digit.
Yet for zero divisor, the reminder is always no less than the divisor.
In this case, MSB of the reminder will be treat as `sign bit' in the succeeding iteration.
Consequently, for zero divisor, the final output quotient of Alg.(\ref{alg_divider}) is
$|1\Bar{a}_{N-1}\cdots \Bar{a}_{1}\rangle$, and the reminder is $|0\cdots01\rangle$.
This result can be problematic.
As positive divisors might also lead to the same output quotient and reminder, it is of great difficulty to clarify the normal divisions and the divisions by zero. 

To address this issue, here we present a tricky solution.
In Alg.(\ref{alg_divider}), the input dividend is a $N$ bit integer $(a_{N-1}\cdots a_0)_{bin}$.
We can also treat the input as a $N+1$ bit integer $(0a_{N-1}\cdots a_0)_{bin}$, where there is an additional 0 as MSB.
Thus, the positive divisors always lead to output quotient as a $N+1$ bit integer, $(0q_{N-1}\cdots q_0)_{bin}$.
As the MSB of the input dividend is 0, the MSB of quotient is also 0.
On the contrary, for zero divisor, the output quotient is $(10\Bar{a}_{N-1}\cdots \Bar{a}_{1})_{bin}$.
To implement this tricky improvement, we only need to delete the first `while' loop in Alg.(\ref{alg_divider}), and set $n\gets N$ before the second `while' loop.
By this mean, normal divisions lead to output quotient with MSB 0, whereas problematic divisions lead to quotient with MSB 1.
Thus, we can distinguish the problematic outputs by testing the MSB.

\section{Discussions}
\label{sec_discussion}

For simplicity, in Sec.(\ref{alg_adder},\ref{sec_subtractor_main},\ref{sec_multiplier},\ref{sec_divider})we always assume that the inputs are at some certain states in the computational basis.
In this section we will demonstrate that the proposed QALUs also support superposition states as input.

Revisit the simplest unit $\hat{P}_{I}$, the quantum circuit is as depicted in Fig.(\ref{fig_qadder_1}), and the qubits connectivity is presented in Fig.(\ref{fig_adder_qubit}a).
The matrix description of $\hat{P}_{I}$ can be given as
\begin{equation}
    \hat{P}_{I}=
    \begin{pmatrix}
    & \alpha_- & 0 & & && &&\\
    & 0 & \alpha_- & & && &&\\
    & && 0&\beta_-& &&&&\\
    & && \alpha_+&0& &&&&\\
    & &&&& 0&\beta_-& &&\\
    & &&&& \alpha_+&0& &&\\
    & &&&&&& \beta_+ &0 &\\
    & &&&&& &0& \beta_+ &
    \end{pmatrix}
    \label{eq_matrix_p1}
\end{equation}
where $\alpha$ and $\beta$ are $4\times 4$ matrices,
\begin{equation}
    \alpha_{\pm}=
    \begin{pmatrix}
        &1 &0&0&0 &\\
        &0 &0&\frac{1\pm i}{2}&\frac{1\mp i}{2} &\\
        &0 &1&0&0 &\\
        &0 &0&\frac{1\mp i}{2}&\frac{1\pm i}{2} &
    \end{pmatrix}
    \label{eq_alpha}
\end{equation}
and
\begin{equation}
    \beta_{\pm}=
    \begin{pmatrix}
        &0 &1&0&0 &\\
        &0 &0&\frac{1\pm i}{2}&\frac{1\mp i}{2} &\\
        &1 &0&0&0 &\\
        &0 &0&\frac{1\mp i}{2}&\frac{1\pm i}{2} &
    \end{pmatrix}
    \label{eq_beta}
\end{equation}
For simplicity, here the matrix is given in the computational basis.
The computational basis are 5 binary digits, corresponding to qubits $A, B, C, C', C''$, from MSB to LSB, as depicted in Fig.(\ref{fig_qadder_1}).
According to Eq.(\ref{eq_matrix_p1}), $\hat{P}_I$ does not change the quantum state of $A$, $B$ (corresponding to the first two digits).
Moreover, if the two input terms are different, one at state $|0\rangle$ while the other at state $|1\rangle$, then $\hat{P}_I$ will flip the state of qubit $C$, which represents the output sum $s$.
The last two digits correspond to the quantum state of qubits $C'$, $C''$.
As discussed in Sec.(\ref{sec_adder}), qubits $C'$, $C''$ are both initialized at state $|0\rangle$.
We have $\alpha_{\pm}|00\rangle=|00\rangle$, and $\beta_{\pm}|00\rangle=|10\rangle$.
In other words, $\alpha$ keeps the input $|00\rangle$ as $|00\rangle$, and $\beta$ flips the input $|00\rangle$ to $|10\rangle$.

Recalling the one-bit quantum full adder with Tofolli gates, as depicted in Fig.(\ref{qadder_1}), the matrix description is
$$
\begin{pmatrix}
    & I & 0 & & && &&\\
    & 0 & I & & && &&\\
    & && 0&X& &&&&\\
    & && I&0& &&&&\\
    & &&&& 0&X& &&\\
    & &&&& I&0& &&\\
    & &&&&&& X &0 &\\
    & &&&&& &0& X &
    \end{pmatrix}
    \label{eq_matrix_pori}
$$
the matrix is also written in the computational basis, corresponding to qubits $A, B, C, C'$, and qubit $C'$ is initialized at state $|0\rangle$.
These two operations lead to same output $s$ and $c_{out}$, when qubits $C', C''$ are prepared at state $|00\rangle$.

According to Eq.(\ref{eq_matrix_p1}), the proposed quantum one-bit full adder $\hat{P}_I$ supports superposition or entanglement in inputs.
Assuming that the input is prepared at a superposition state $\sum_{j,k,l=0,1}\gamma_{j,k,l}|j,k,l,0,0\rangle$, where $\sum_{j,k,l=0,1}|\gamma_{j,k,l}|^2=1$.
Then we have the output is also a superposition state,
\begin{equation}
    \begin{split}
        &\hat{P}_{I}(A, B, C, C', C'')
    \sum_{j,k,l=0,1}\gamma_{j,k,l}|j,k,l,0,0\rangle
    \\
    =&\sum_{j,k,l=0,1}\gamma_{j,k,l}|j,k,s_{j,k,l},c_{j,k,l},0\rangle
    \end{split}
    \label{eq_p1_super}
\end{equation}
where $s_{j,k,l}=j\oplus k\oplus l$ is the output sum, and $c_{j,k,l}=(j\cdot k)+(l\cdot(j\oplus k))$ is the output carry.
The quantum one-bit full adder $\hat{P}_I$ does not change the values of $\gamma$.
Instead, $\hat{P}_I$ converts the input from one state in the computational basis to another state in the computational basis.
Eq.(\ref{eq_p1_super}) can lead to some interesting results.
For example, if the input is initialized at Bell state\cite{nielsen2010quantum} (the two input terms are entangled), then we have
\begin{equation}
    \begin{split}
        &\hat{P}_{I}\left(\frac{1}{\sqrt{2}}(|0,0,0,0,0\rangle+|1,1,0,0,0\rangle)
        \right)
        \\
        =&
        \frac{1}{\sqrt{2}}(|0,0,0,0,0\rangle+|1,1,0,1,0\rangle)
    \end{split}
\end{equation}
where for simplicity, $\hat{P}_{I}(A, B, C, C', C'')$ is abbreviated as $\hat{P}_{I}$.
By this mean, the input Bell state is converted to the GHZ state\cite{greenberger1989going, mermin1990quantum}, where qubits $A,B,C'$ are entangled.

Then study the another quantum one-bit full adder, $\hat{P}_{II}$.
The quantum circuit of $\hat{P}_{II}$ is depicted in Fig.(\ref{fig_qadder_2}).
The matrix description of $\hat{P}_{II}$ is exactly the same as $\hat{P}_{I}$.

As for ${P}_{III}$, the fundamental units are $U_C$ and $U_S$, as depicted in Fig.(\ref{fig_qadder_sp1}).
$U_S$ is formed by two CNOT gates, whereas the implementation of $U_C$ is more intricate.
The matrix description of $U_C$ is
\begin{equation}
    U_{C}=
    \begin{pmatrix}
    & \alpha_- & 0 & & && &&\\
    & 0 & \alpha_- & & && &&\\
    & & &\alpha_+&0 & &&&&\\
    & & &0&\beta_- & &&&&\\
    &&& & &\alpha_+&0 & &&\\
    &&& & &0&\beta_- & &&\\
    & &&&&&& \beta_+ &0 &\\
    & &&&&& &0& \beta_+ &
    \end{pmatrix}
    \label{eq_matrix_uc}
\end{equation}
The matrix is expanded in the computational basis.
The computational basis are 5 binary digits, corresponding to qubits $A_n,B_n,C_{n},C_{n+1},C_{n+2}$, from MSB to LSB.
Qubits $C_{n},C_{n+1}$ are prepared at state $|00\rangle$.
$U_C$ does never change the state of qubits $A_n,B_n,C_{n}$.
Thus, the first three digits in the quantum state do not change after applying $U_C$.
The fourth digit, corresponding to the state of qubit $C_{n+1}$, which re[resents the output carry.
The output carry is 1 if there are two or three 1 in the inputs (two input terms, and one input carry), corresponding to the first three digits.
Otherwise, the output carry is 0.
Therefore, if the first three digits are 011, 101, 110, or 111, $U_C$ flips the fourth digit from state $|0\rangle$ to $|1\rangle$ (corresponding to the $\beta$ operations in Eq.(\ref{eq_matrix_uc})), otherwise $U_C$ keeps the fourth digit at state $|0\rangle$ (corresponding to the $\alpha$ operations operations in Eq.(\ref{eq_matrix_uc})).
The last digit, corresponds to the quantum state of ancilla qubit $C_{n+2}$.
$C_{n+2}$ is initialized at state $|0\rangle$, and the output state of $C_{n+2}$ is also $|0\rangle$.
For input superposition state, we have
\begin{equation}
    \begin{split}
        &U_C
    \sum_{j,k,l=0,1}\gamma_{j,k,l}|j,k,l,0,0\rangle
    \\
    =&\sum_{j,k,l=0,1}\gamma_{j,k,l}|j,k,l,c_{j,k,l},0\rangle
    \end{split}
    \label{eq_uc_super}
\end{equation}
where $U_C(A_n,B_n,C_{n},C_{n+1},C_{n+2})$ is abbreviated as $U_C$, $c_{j,k,l}=(j\cdot k)+(l\cdot(j\oplus k))$ is the output carry.

The proposed quantum adders $P_I$, $P_{II}$ and $P_{III}$ are formed as a cascade of these fundamental units.
The quantum adders also support superposition states as input. 
For $P_I$, we have
\begin{widetext}
\begin{equation}
    \begin{split}
        &P_{I}(A_{N-1},\cdots A_0; C_{N+1},C_N,\cdots,C_0; B_{N-1},\cdots, B_0)
        \sum_{x=0}^{2^N-1}\sum_{y=0}^{2^N-1}
    \gamma_{xy}|x_{N-1}\cdots x_0;00\cdots0;y_{N-1}\cdots y_0\rangle
    \\=&
    \sum_{i=0}^{2^N-1}\sum_{j=0}^{2^N-1}
    \gamma_{ij}|x_{N-1}\cdots x_0;0s_N\cdots s_0; y_{N-1}\cdots y_0\rangle
    \end{split}
\end{equation}
\end{widetext}
where $\sum_{i,j=0}^{2^N-1}|\gamma_{ij}|^2=1$.
For simplicity, the two input integers are denoted as $x,y=0,1,\cdots,2^N-1$, corresponding to the N-bit binary integers $(x_{N-1}\cdots x_0)_{bin}$, $(y_{N-1}\cdots y_0)_{bin}$, and $x_n$, $y_n$ are the $n^{th}$ digits.
The sum is a $N+1$ bit binary integer $(s_{N}\cdots s_0)_{bin}$.
As depicted in Fig.(\ref{fig_adder_qubit}c), there are $N+2$ qubits in the $C$ column, whereas the output sum is a $N+1$ bit binary integer.
Thus, there is an additional 0 in the output.
For $P_{II}$, we have
\begin{widetext}
    \begin{equation}
    \begin{split}
        &P_{II}(A_{N-1},\cdots A_0; B_{N-1},\cdots, B_0; C_{N+1},C_N,\cdots,C_0)
        \sum_{x=0}^{2^N-1}\sum_{y=0}^{2^N-1}
    \gamma_{xy}|x_{N-1}\cdots x_0;y_{N-1}\cdots y_0; 00\cdots0\rangle
    \\=&
    \sum_{i=0}^{2^N-1}\sum_{j=0}^{2^N-1}
    \gamma_{ij}|x_{N-1}\cdots x_0;y_{N-1}\cdots y_0; 0s_N\cdots s_0\rangle
    \end{split}
\end{equation}
\end{widetext}
In $P_I$, $P_{II}$, the output sum is stored in a standalone column.
$P_{III}$ is a little different, as the output summary is stored in the $B$ column, which represents a input term at beginning.
Thus we have
\begin{widetext}
    \begin{equation}
    \begin{split}
        &P_{III}(A_{N-1},\cdots A_0; B_N,B_{N-1},\cdots, B_0; C_N,C_{N-1}\cdots,C_0)
        \sum_{x=0}^{2^N-1}\sum_{y=0}^{2^N-1}
    \gamma_{xy}|x_{N-1}\cdots x_0;0y_{N-1}\cdots y_0; 00\cdots0\rangle
    \\=&
    \sum_{i=0}^{2^N-1}\sum_{j=0}^{2^N-1}
    \gamma_{ij}|x_{N-1}\cdots x_0;s_Ns_{N-1}\cdots s_0;00\cdots 0\rangle
    \end{split}
    \label{eq_p3_super}
\end{equation}
\end{widetext}

Other arithmetic circuits also supports superposition states as input.
In Sec.(\ref{sec_subtractor_main}), we present the $U_\pm$ to calculate the two's complement description of a given signed binary integer, and the quantum circuit of binary subtractor.
$U_\pm$ contains three main steps: the first step is to flip all bits of the absolute (by applying Pauli-X gate), then add 1 to the inverted result, ignoring the overflow (by applying $\Tilde{P}_{+1}$), and the last step is to reset the ancilla qubits (by applying a series of CNOT gates).
The key operation $\Tilde{P}_{+1}$, can be regard as a version of $P_{III}$, where one input term is set as 1, and the overflow is ignored.
For $U_\pm$, we have
\begin{equation}
    \begin{split}
        &U_{\pm}\sum_{x=0}^{2^{N+1}-1}\gamma_{x}|x_N,x_{N-1}\cdots x_0; 0\cdots 0\rangle
    \\=&\sum_{x=0}^{2^{N+1}-1}\gamma_{x}
    |x_N,\Tilde{x}_{N-1}\cdots\Tilde{x}_0; 0\cdots 0\rangle
    \end{split}
    \label{eq_upm_super}
\end{equation}
where we have $\sum_{x=0}^{2^{N+1}-1}|\gamma_{x}|^=1$, and $U_{\pm}(A_N,\cdots,A_0;C_N,\cdots, C_0)$ is abbreviated as $U_{\pm}$.
The binary description of integer $x$ is a N+1 bit binary integer $(x_{N}x_{N-1}\cdots x_0)_{bin}$, and we use this binary string to represent a signed integer, where $x_{N}$ is the sign bit.
In this context, $(x_{N-1}\cdots x_0)_{bin}$ is the absolute, and $(\Tilde{x}_{N-1}\cdots\Tilde{x}_0)$ is the two's complement representation.

With $U_{\pm}$ and the proposed quantum adders, we are able to implement subtraction on quantum computers.
Firstly apply $U_{\pm}$ and obtain the two's complement representation of the subtrahend, and then apply $P_{III}$ to sum them up.
By this mean, the difference is obtained (in two's complement representation).
Nevertheless, $U_{\pm}$, and the proposed quantum adders, especially $P_{III}$ and $\Tilde{P}_{III}$, constitute the core of the proposed quantum multiplier and divider.
The quantum multiplier based on long multiplication, as presented in Alg.(\ref{alg_qmul}), is implemented by iteratively applying $P_{III}$ and $U_{\pm}$, along with shifting from LSB to MSB.
Similarly, the quantum divider, as presented in Alg.(\ref{alg_divider}), is implemented by iteratively applying $\Tilde{P}_{III}$ and $U_{\pm}$, along with shifting from MSB to LSB.
Like $U_{\pm}$ and $P_{III}$, the proposed quantum multiplier and divider also support superposition states as input.

\section{Conclusion and Outlook}
\label{sec_conclusion}

In this paper, we propose feasible quantum adder, subtractor, multiplier and divider for near-term quantum computers, where the qubits are assigned in two-dimensional arrays.
In Sec.(\ref{sec_adder}), we present feasible and scalable quantum full adders $P_I$, $P_{II}$ for near-term quantum computers, where the input terms are mapped in two columns of qubits, and the output sum is stored in another column of qubits.
Moreover, a special quantum adder $P_{III}$ is introduced for iterative additions, where one input column is used to store the output sum.
In Sec.(\ref{sec_subtractor_main}), we demonstrate the process for obtaining the two's complement representation of a signed binary integer using operation $U_\pm$.
The two's complement representation enables subtraction using the proposed quantum adders. 
Operations $P_{III}$ and $U_\pm$ form the core of the proposed quantum multiplier and divider, as discussed in Sec.(\ref{sec_multiplier}) and Sec.(\ref{sec_divider}).
The processes of long multiplication and division are broken down into iterative shifting, additions, and subtractions.
Thus, by iteratively applying $P_{III}$ and $U_{\pm}$ and appropriately shifting temporary results, we can perform multiplications and divisions on quantum computers.

The key features of the proposed QALUs are as follows.

\noindent
{\bf 1.} Feasibility.
The proposed QALUs can be implemented using only Pauli-X gates, CNOT gates and $C\sqrt{X}$ (CSX) gates, and all of the two-qubit gates perform on nearest neighbor qubits.
Therefore, the proposed QALUs are feasible for implementation on near-term quantum computers with qubits arranged in two-dimensional arrays.

\noindent
{\bf 2.} Efficient use of ancilla qubits.
In the proposed QALUs, the ancilla qubits are all initialized at state $|0\rangle$, and are reset to state $|0\rangle$ by the end.
This allows for the reuse of ancilla qubits in subsequent operations. Additionally, the circuits avoid the need for extra qubits to store intermediate results, even for complex operations like multiplication and division.

\noindent
{\bf 3.} Support for superposition states.
As discussed in Sec.(\ref{sec_discussion}), the QALUs are designed to handle superposition states as input.
They can process qubits that are in a quantum superposition of multiple states simultaneously. The output of these circuits is also a superposition state, transforming the input from one basis state to another within the computational basis.

We propose specific areas where further investigation could expand upon our work, addressing the limitations identified and exploring new questions that have emerged.
The potential directions for future research are as follows. 

{\bf The minimum requirement of qubit connectivity.}
The QALUs are designed for quantum computers with a square qubit lattice layout. 
However, not all advanced quantum computers have this architecture.
It is crucial to study the minimum necessary qubit connectivity and determine if the proposed QALUs can function on quantum computers with different architectures, such as IBM's quantum computers.

{\bf Readout of superposition outputs and potential quantum speedup.}
In Sec.(\ref{sec_discussion}), we demonstrate that the proposed QALUs support superposition states as inputs and produce corresponding superposition states as outputs. 
A critical question is how to efficiently read out and utilize the output superposition states.

{\bf Compatibility with quantum error correction techniques.}
This paper focuses on the feasibility of the QALUs and does not extensively address noise and errors.
Advanced quantum error correction techniques are essential for achieving accurate results.
Optimizing the proposed QALUs with the assistance of error correction techniques is a meaningful avenue for further research.

{\bf From integer to float.}
In classical computing, floating-point arithmetic (FP) represents subsets of real numbers using an integer with a fixed precision, called the significand, scaled by an integer exponent of a fixed base\cite{muller2018handbook}.
Recently, quantum circuits for floating-point addition and multiplication have been proposed, which require long-range two-qubit gates and Toffoli gates\cite{haener2018quantum, seidel2022efficient}, making their implementation challenging on near-term quantum computers. 
This paper focuses on QALUs for integers.
An interesting open question is whether the proposed QALUs can be generalized to the FP scheme using Pauli-X gates, CNOT gates, and $C\sqrt{X}$ that operate on neighboring qubits.
If the proposed QALUs can be extended to FP arithmetic, they could approximate functions that are difficult to compute, such as exponential functions and logarithms.

In conclusion, we proposes elementary quantum arithmetic logic circuits (QALUs) for performing addition, subtraction, multiplication, and division on near-term quantum computers.
These operations are designed to be scalable and feasible for implementation on near term quantum hardware, where qubits are arranged in two-dimensional arrays.
The proposed QALUs can be implemented using only Pauli-X gates, CNOT gates and $C\sqrt{X}$ (CSX) gates, and all of the two-qubit gates perform on nearest neighbor qubits.
Besides, another key advantage of our approach is the efficient use of ancilla qubits.
Ancilla qubits are initialized to the state 
$|0\rangle$ and are reset to $|0\rangle$ at the end of computations, allowing them to be reused in subsequent operations.
This approach reduces the overall number of qubits required, making the circuits more practical for implementation on current quantum hardware, which typically has limited qubit counts.
Furthermore, the proposed QALUs support superposition states as input and produce output in superposition states as well. This capability is essential for maintaining the quantum nature of computations and for leveraging the parallelism inherent in quantum computing. The ability to handle superposition states makes these circuits suitable for integration into more complex quantum algorithms.
In summary, our work provides feasible, resource-efficient, and scalable methods for performing basic arithmetic operations on near-term quantum computers.
The proposed QALUs pave the way toward more complex quantum algorithms and computations.

\section{Acknowledgement}
J.L gratefully acknowledges funding by National Natural Science Foundation of China (NSFC) under Grant No.12305012.
\bibliography{ref}
\end{document}